\documentclass[aps,preprint,prd,eqsecnum,nofootinbib,showpacs,12pt]{revtex4}

\usepackage{epsfig}
\usepackage{bm}

\newcommand{\bq}{\begin{equation}}
\newcommand{\be}{\begin{equation}}
\newcommand{\eq}{\end{equation}}
\newcommand{\ee}{\end{equation}}
\newcommand{\bqa}{\begin{eqnarray}}
\newcommand{\ba}{\begin{eqnarray}}
\newcommand{\eqa}{\end{eqnarray}}
\newcommand{\ea}{\end{eqnarray}}

\begin{document}
\pagestyle{empty}

\title{
{\vspace{-1.2em} \parbox{\hsize}{\hbox to \hsize
{\hss \normalsize\rm PM/03-04}}} \\
Sudakov expansions at one loop and beyond\\
for charged scalar and fermion pair production in SUSY models at future Linear Colliders
\footnote{Partially supported by EU contract HPRN-CT-2000-00149}}

\author{Matteo Beccaria}
\email{Matteo.Beccaria@le.infn.it}
\affiliation{INFN, Sezione di Lecce, and
Dipartimento di Fisica dell'Universit\`a di Lecce,
Via Arnesano, ex Collegio Fiorini, I-73100 Lecce, Italy}

\author{Michael Melles} 
\email{michael.melles@psi.ch}
\affiliation{Paul Scherrer Institute (PSI), CH-5232 Villigen, Switzwerland}

\author{Fernand M. Renard}
\email{renard@lpm.univ-montp2.fr}
\affiliation{ Physique Math\'{e}matique et Th\'{e}orique, UMR 5825\\
Universit\'{e} Montpellier II,  F-34095 Montpellier Cedex 5.}

\author{Salvatore Trimarchi}
\email{trimarchi@ts.infn.it}
\author{Claudio Verzegnassi}
\email{verzegnassi@ts.infn.it}
\affiliation{Dipartimento di Fisica Teorica, Universit\`a di Trieste, \\
Strada Costiera  14, Miramare (Trieste) \\
INFN, Sezione di Trieste
}

\begin{abstract}
We consider the high energy behaviour of the amplitudes for pair
production of charged leptons, quarks, Higgs bosons, sleptons,
squarks and charginos at lepton colliders. We give the
general expressions of the leading quadratic and subleading linear
logarithms that appear at the one loop level, and derive the
corresponding resummed expansions to \underline{subleading}
logarithmic order accuracy. Under the assumption of a relatively light
SUSY scenario and choosing the MSSM as a specific model, we compare
the predictions of the one-loop and of the resummed expansions at
variable energy. We show that the two predictions are
very close in the one TeV regime, but drastically differ in
the few ($2,3$) TeV range.
\end{abstract}


\maketitle

\pagestyle{plain}

\section{Introduction}

It is by now well-known that the electroweak radiative corrections to
Standard Model pair production processes increase strongly with the
center of mass energy $\sqrt{s}$ at the one-loop level. This is due to
the presence of large double and single logarithms $\simeq
{\alpha\over\pi}\log^2{s\over M^2_W}$, ${\alpha\over\pi}\log{s\over M^2_W}$
\cite{sud1,sud2,dp1,dp2,log}.
In the TeV range such terms reach the several percent level,
which should be easily observable (and measurable) at future lepton
colliders \cite{LC,CLIC}.
Actually, for energies beyond the few TeV range, the
numerical size of their effect begins to be too large \cite{log}
(beyond the
relative ten percent level), and the validity of the simple one-loop
approximation must be seriously questioned. This has led a number of
theoretical groups to propose resummation prescriptions
\cite{resum,flmm,kmps,habil}.
Without entering the details of the different calculations,
we shall accept the conclusion that,
for \underline{massive} final pairs, an agreement
seems to exist that a full resummation can be given, but only to
\underline{subleading} logarithmic level.\\
In the case of supersymmetric extensions of the SM, similar studies
have been recently performed for the processes of production of
fermion pairs at the one-loop level \cite{brv1}
and of scalar pairs \cite{bmrv} in the
MSSM. In ref.\cite{bmrv} both the one-loop approximation
and the resummed
expansions have been computed. Under the qualitative assumption of a
relatively light SUSY scenario,  with all the relevant masses of
the process "adequately" smaller than e.g. one TeV, it has been shown
that the two approximations are "essentially" (i.e. at the expected
relative one percent level of experimental accuracy) identical in the
$\simeq1$ TeV region (final possible reach of the proposed LC
\cite{LC}), but
differ drastically in the few ($2,3$) TeV range (aim of the proposed
CLIC \cite{CLIC}). Thus, for the spin zero production case, 
\underline{to subleading logarithmic accuracy}, a simple
one-loop approximation would seem definitely inadequate at CLIC
energies, but valid  in
the LC regime.\par
The aim of this paper is that of performing a general investigation,
analogous to that already carried through for scalar production,
to include also the processes of charged spin one-half (fermions,
charginos) pair production.  For all these cases we shall
give both the one-loop level and the resummed logarithmic expansions
to subleading logarithmic accuracy. Under the assumption of a
relatively light SUSY scenario, we shall then compare the two
approximations at variable c.m. energies for a number of experimental
observables. We shall show that the same conclusions obtained for
scalar production are apparently valid for the extended case, making
the validity to subleading logarithmic order of a one-loop approximation 
for spin zero and one-half charged pair production in the MSSM
 at future colliders \underline{to only depend on the chosen
c.m. energy}, \underline{and not on the specific process}. Our
analysis has been performed and our conclusions have been drawn in the
particular case of the MSSM, but they could be easily adapted or
modified to treat different supersymmetric models, although we believe
that this is beyond the purposes of this first, necessarily
preliminary work.\par
Technically speaking, the plan of this paper is the following: Section
2 contains the one-loop expansion of all the considered processes. The
resummed expansions are shown in Section 3, and a comparison between
the two approximations is given in Section 4. A final discussion and
some possible conclusions are given in Section 5. The
definitions of the observables for final fermionic or scalar
pairs are given in Appendix A. Appendices B and C contain a summary
of several long (but necessary) analytic formulae
for the asymptotic expansions at the one loop level and with
resummation.

\section{Asymptotic logarithmic expansion at one-loop}

\subsection{Generalities}

As illustrated and discussed in several
previous papers \cite{brv1,bmrv,brv4},
at the one-loop level, the logarithmic terms appearing in $e^+e^-$
annihilation processes can be
separated into three categories, Renormalization Group (RG) terms,
Universal terms and
Non Universal (angular dependent) terms, so that
the invariant scattering amplitude for the pair production process
\be
e^+_{\alpha} e^-_{\alpha} \longrightarrow {f}_{1}{f}_{2}
\ee
\noindent
($\alpha$ representing the electron chirality)
can be written in the following form
\be
A^{\rm 1\ loop} = A^{\rm RG}+ A^{\rm univ}+ A^{\rm non~univ}
\label{a1l}\ee

\underline{The RG contribution} represents the linear logarithms
\cite{DS} generated by the "running" of the gauge
coupling constants, that are known and
calculable in a straightforward way. It is
obtained by introducing in $A^{\rm Born}$
the running $SU(2)\times U(1)$ couplings $g,g'$
according to the asymptotic MSSM
${\tilde \beta}_0,~{\tilde \beta}_0^\prime$ functions:
 \begin{eqnarray}
{\tilde \beta}_0&=& \frac{3}{4} C_A- \frac{n_g}{2}-\frac{n_h}{8} \;\;,\;
 {\tilde \beta}_0^\prime=-\frac{5}{6}n_g-\frac{n_h}{8} \label{eq:bMSSM} \\
 g^2(s) &=& \frac{g^2(\mu^2)}{1+{\tilde \beta}_0 \frac{g^2
 (\mu^2)}{4\pi^2}
 \log \frac{s}{\mu^2}} \;\;,\;
 {g^\prime}^2 (s) = \frac{{g^\prime}^2 (\mu^2)}{1+{\tilde \beta}^\prime_0
 \frac{{g^\prime}^2 (\mu^2)}{4\pi^2}
 \log \frac{s}{\mu^2}} \label{aRG}
 \end{eqnarray}
 where $C_A=2$, $n_g=3$, $n_h=2$ and $g\sin\vartheta_{\rm w} = 
g'\cos\vartheta_{\rm w} = e$.

At one loop, this quite general procedure gives single logarithms
that do not factorize, but can be obtained from $A^{\rm Born}$ as:

\begin{equation}
 A^{\rm RG}=-{1\over4\pi^2}~\left(g^4{\tilde \beta}_0{dA^{\rm Born}
\over dg^2}+
~g^{'4}{\tilde \beta}_0^\prime{dA^{\rm Born}
\over dg^{'2}}~\right)\  \log{s\over\mu^2}
\label{RGder}\end{equation}
\noindent
where $\mu$ is a reference scale defining the numerical values
of $g,g'$.\\

\underline{The universal contribution}
can be of quadratic and of linear kind and in a covariant gauge 
is produced by diagrams of vertex (initial and final triangles)
and of box type;
it only depends on the quantum
numbers of the initial
$e^+,e^-$ and final $f_1,f_2$ lines and can be written as:

\be
A^{\rm univ} = A^{\rm Born}.~\left(~c^{\rm in}_{\alpha}+c^{\rm fin}~\right)
\label{auniv}\ee
\noindent
with the correction to the initial $e^+e^-$ lines
\bqa
c^{\rm in}_{\alpha} &=& \frac{1}{16\pi^2}\left( g^2
I_{e^-_{\alpha}}(I_{e^-_{\alpha}}+1)+
  {g^\prime}^2\frac{Y^2_{e^-_{\alpha}}}{4} \right)
\left( 2 \log \frac{s}{M_V^2} -\log^2\frac{s}{M_V^2}\right)
\label{cin}\eqa
\noindent
where $M_V$ is a common gauge boson mass (we use $M_Z\simeq M_W$
for simplicity and we have separated the infrared part of the photon
contribution, keeping here only the ultraviolet one by putting
$M_{\gamma}=M_Z$), $\alpha$ is the chirality and $I,Y$ the isospin
and hypercharge of the initial electron $e^{-}$. The expression
for $c^{\rm fin}$ will be given separately
for each of the examples listed below.\\

\underline{The non universal (angular dependent) contribution}
can be written
\be
A^{\rm non ~univ} = A^{\rm Born} \cdot c^{\rm ang}
\label{anu}\ee
\noindent
It only consists in residual terms arising
from the quadratic logarithms
$\log^2t,~\log^2u$ (from which the $\log^2s$ part has been
subtracted and put in the universal contribution)
generated by box diagrams
containing $W,Z,\gamma$
gauge boson internal lines where
$t=-\frac{s}{2} \left( 1-\cos \vartheta \right)$
and $u=-\frac{s}{2} \left( 1+\cos
\vartheta \right)$,~
$\vartheta$ being the scattering angle.
There are only few such diagrams and they have been all
explicitely computed.\\

To subleading logarithmic accuracy,
these three types of contributions have been calculated exactly.
We shall now summarize, trying to make the review as short
as possible but reasonably complete and self-contained,
the results of our effort
for the charged final pairs of the MSSM, including those
for scalar production and fermion production that were
already given in other papers.
The list of considered
final states is now given in the following subsections.

\subsection{Chiral Lepton or Quark Pair ${\overline f}_{\beta}
f_{\beta}$}

The Born amplitude is given in Appendix B1, using the 
notations given in Appendix A; for more details
see previous papers \cite{brv1}.
For a final fermion pair with chirality $\beta$,
the logarithmic part of the
one loop amplitude (\ref{a1l}) is obtained by adding
eq.(\ref{RGder}),(\ref{auniv}),(\ref{anu})
with:

\be
c^{\rm fin}_{\beta}=c^{\rm fin~ gauge }_{\beta}
+c^{\rm fin~ Yukawa}_{\beta}
\ee
\begin{eqnarray}
\label{fermionstart}
c^{\rm fin~ gauge}_{\beta} &=& \frac{1}{16\pi^2}\left( g^2 
I_{f_\beta}(I_{f_\beta}+1)+
  {g^\prime}^2\frac{Y^2_{f_\beta}}{4} \right)
\left( 2 \log \frac{s}{M_V^2} -\log^2\frac{s}{M_V^2}\right) \\
c^{\rm fin~Yukawa}_{\beta} &=&  - \frac{ g^2}{16 \pi^2} \left(
\frac{1+\delta_{\beta,{\rm R}}}{2}
\frac{{\hat m}
 ^2_f}{M^2_V} + \delta_{\beta,{\rm L}}
 \frac{{\hat m}^2_{f^\prime}}{2 M^2_V} \right)
 \log \frac{s}{M_S^2}
\end{eqnarray}

and
\bqa
c^{\rm ang}_{\alpha\beta} &=& -\frac{g^2}{16\pi^2} \log \frac{s}{M^2} \left[
\left( \tan^2
\vartheta_{\rm w}
Y_{e^-_{\alpha}} Y_{
f_\beta}
+ 4 I^3_{e^-_{\alpha}} I^3_{f_\beta} \right) \log \frac{t}{u} \right. \nonumber
\\ &&
 \left. + \frac{\delta_{\alpha, L} \delta_{\beta,L}}{\tan^2 \vartheta_{\rm w}
Y_{e^-_{\alpha}} Y_{
f_\beta} /4
+ I^3_{e^-_{\alpha}} I^3_{f_\beta}} \left( \delta_{d,f} \log \frac{-t}{s} -
\delta_{u,f}
\log \frac{-u}{s} \right)  \right]
\label{fermionend}
\eqa
\noindent
We denote ${\hat m}_f=m_t / \sin \beta$
if $f=t$ and ${\hat m}_f=m_b / \cos \beta$ if $f=b$; $f^\prime$ denotes the
corresponding
isopartner of $f$, and $\beta$ (not to be confused with the
chirality index) is the mixing angle between the vacuum
expectation values of the up and down Higgs chiral superfield (in standard
notation
$\tan\beta = v_u/v_d$).

For particles other than those belonging to the third family
 of quarks/squarks, the Yukawa terms are negligible.

\subsection{Slepton or Squark Pair $  \widetilde f^*_\beta \widetilde
f_\beta$}

The results for this case are quite similar to those valid for fermion
production, see Appendix B2 and ref.\cite{bmrv}.
The logarithmic part of the one loop contribution for the process
\be
e^+_{\alpha} e^-_{\alpha}
 \longrightarrow \widetilde f^*_\beta \widetilde
f_\beta
\ee
($\beta=L,R$) can be written in the same form as for
$e^+_{\alpha} e^-_{\alpha}
 \longrightarrow {\overline f}_{\beta}
f_{\beta}$ producing a
fermion pair with chirality $\beta$,
with the same expressions for $c^{\rm in}_{\alpha}$,
$c^{\rm fin~gauge}_{\beta}$, $c^{\rm fin~Yukawa}_{\beta}$
and $c^{\rm ang}_{\alpha\beta}$.

\subsection{Charged Higgs Bosons $H^{\pm}$ or charged Goldstones
$G^{\pm}$}

In the MSSM the charged Higgs bosons and Goldstones
are produced in pairs through the same diagrams that appear
in the case of sfermion production.
At $M^2_W/s$ accuracy the amplitudes for $G^{+}G^{-}$ production
are equivalent to the physical amplitudes for
longitudinal $W^{+}_LW^{-}_L$ states.
The previous equations,
(\ref{fermionstart}-\ref{fermionend}) which concern the gauge parts
give the correct result for the process
\be
e^+_{\alpha} e^-_{\alpha} \longrightarrow H^+H^-
\ee
provided that $f_\beta$ is replaced by $H^-$ with the
following quantum numbers
\be
Q(H^-)=-1,\qquad I^3(H^-) = -\frac 1 2
\ee
For what concerns the Yukawa part, one has
\begin{eqnarray}
c^{\rm fin~Yukawa} &=&  - 3\frac{ g^2}{32 \pi^2} \left(
\frac{m^2_t\cot^2\beta}{M^2_W}+\frac{m^2_b\tan^2\beta}{M^2_W} \right)
 \log \frac{s}{M_S^2}
\label{yukhh}
\end{eqnarray}

These formulae then  apply also to
\be
e^+_{\alpha} e^-_{\alpha} \longrightarrow G^+G^-
\ee
without any change concerning the Born term and the one loop
gauge terms, but the Yukawa part has to be simply modified with
$m^2_b \tan^2\beta~\to~m^2_b$ and $m^2_t \cot^2\beta~\to~m^2_t$.

\subsection{Charginos $\chi^+_i\chi^-_j$}

Charginos are mixtures of gaugino(Wino) and Higgsino components
selected respectively by the mixing matrix elements $Z^{\pm}_{1i},~
Z^{\pm}_{2i}$ using the notations of ref.\cite{Rosiek}.
The expressions of the Born terms and the one-loop contributions
written in the Appendix B3
reflect clearly the properties of these two types of components.\par

The Born amplitude involves  both Wino components
($Z^{\pm}_{1k}Z^{\pm}_{1j}$)
and Higgsino components ($Z_{2i}^{\pm}Z_{2j}^{\pm}$)
produced in $s$-channel through $\gamma,~Z$ formation, but it only
involves
Wino components produced through sneutrino exchange in the $u$-channel
(the contribution of the Higgsino component vanishes like
the electron mass).\par

The RG amplitude is computed using Eq.(\ref{RGder}). We have written
separately its Higgsino part. Its Wino part is regrouped with the other
one loop Wino contributions because of remarkable cancellation
properties explained below.\par

The universal terms $c^{\rm fin}$ generated
by the $\chi^+_i\chi^-_j$ lines contain Higgsino and Wino components
that factorize the corresponding Born term components. We have
computed them both through one loop diagrams in the 't
Hooft-Feynman gauge and through the splitting function formalism~\cite{Splitting}
with addition of Parameter Renormalization terms obtaining
an agreement that we consider, given the not simple structure of the 
related formulae, encouraging. We observed the
following properties. The $s$-channel terms contain
universal corrections from both Higgsino and
Wino components; their Wino part can be identified
with the sum of a $\chi^{\pm}$ splitting function contribution
and of a Wino Parameter Renormalization term (the RG Wino part
computed through eq.(\ref{RGder}),
see details in Appendix B3);
its Higgsino part contains both a "gauge" and a "Yukawa"
part. The $u$-channel terms only contain a Wino part
also identifiable with the sum of a $\chi^{\pm}$ splitting function
contribution and of a Wino Parameter Renormalization term.
These properties of the Wino and of the Higgsino parts
are similar to those observed in
the cases of $W^+W^-$ and $H^+H^-$ production, respectively; in
particular at one loop
we get only a DL term (the SL terms cancel) for the
universal "gauge" parts both in $W^+W^-$ and in Wino pair
production.\par

The non universal (angular dependent) terms $c^{\rm ang}$
arise from residual terms of $\gamma\gamma$,
$\gamma Z$, $ZZ$ and $WW$ boxes in the $s$ channel and of
single $\gamma$, $Z$, and $W$ boxes
in the sneutrino exchange channel leading to both $1/u$ and $1/t$
terms (see Appendix B3).\par

A point that must be noticed is the fact that an important difference
exists between the chargino pair production and the previous
considered processes. This is due to the fact that, in the chargino
case, the Born approximation already contains typical SUSY parameters
in the mixing coefficients. Being by definition "bare" quantities, a
suitable extra renormalization is required. This is a well-known fact,
already discussed in previous papers (see e.g. \cite{Hollik}), and the
choice of a convenient renormalization scheme is essential. In our
case, at the chosen level of logarithmic accuracy, we can neglect
this complication since the difference between the bare parameters and
the physical ones will be in any case a constant term of order
$\alpha$, and will not find place in our expansion (it should be,
though, properly retained in a more complete next-to next to leading
order- analysis). This means that in our one-loop expansions one can
systematically assume, for the values of these parameters, those of
the corresponding physical ones in the suitable renormalization scheme
that has been adopted (in our case, we are using systematically the 
minimal reduction scheme).

\section{Resummation of subleading logarithms in general
susy-processes}

The size of the one loop SUSY Sudakov corrections, investigated
in Refs.\cite{brv1,bmrv,brv3,brv4},
indicates that at TeV energies one must in principle also
include the higher
order contributions as mentioned in the introduction.
While for SM processes a lot of attention has been devoted
to this particular problem in Refs.
\cite{flmm,brv1,m1,m2,m3,m4,kps,kmps,dp1,dp2,ccc1,hkk,bw},
for SUSY Sudakov logarithms only
higher order corrections to scalar production in $e^+e^-$
collisions are known
\cite{bmrv}.

We therefore discuss the situation in the MSSM for final
fermion production
in some more detail in order to clarify the arguments.
As in Ref. \cite{bmrv}
we assume a ``light mass'' SUSY scenario (with the SUSY mass scale 
$m_s \sim m_t \sim m_H \equiv M$ to logarithmic accuracy) for all particles
involved in the loop corrections.\\

For what concerns  the DL and angular dependent logarithmic corrections, 
since they are only 
mediated  by the exchange of SM gauge bosons, their treatment will be identical 
with that already given in the Standard Model case
\cite{flmm,m5,dmp}, so that their exponentiation will be granted at the same
subleading logarithmic order accuracy, and does not need to be rediscussed here.\\

For the subleading Sudakov corrections of the universal,
i.e. process
independent type, we have now novel contributions in the
MSSM, both from particles
with  electroweak gauge couplings and from particles with coupling of Yukawa
type. Since the former where already discussed in Ref.
\cite{bmrv}, we will
only discuss here the novel Yukawa type Sudakov terms for third
family quarks which are of particular
interest since they contain a strong $\tan \beta$ dependence.
For the purposes of this paper we shall only consider
 the asymptotic  corrections above the scale of electroweak symmetry
breaking since
the mass-gap contributions originate only from QED and are
therefore in
principle known \cite{flmm,habil}. For realistic collider
simulations they must, though, carefully
be included via matching at the weak scale.

In order to establish the ``exponentiation'' of the one
loop Yukawa corrections,
i.e. the fact that we only need to consider the situation depicted
in Fig.~(\ref{fig:exp}),
we need to show that the diagrams of the type shown in
Fig.(\ref{fig:wi}) cancel each-other.

As we argued in Ref. \cite{bmrv,habil,m6,bmrv2,m7}
using the \underline{symmetric} basis,
this feature is ensured by the gauge invariance of
the Yukawa sector. In order to demonstrate the
cancellation in the \underline{physical} basis
let us first consider the case of right handed external top quarks.

In the following we only need to consider soft gauge boson insertions
(with loop momentum $l$) since we want
to generate three large logarithms at the two loop level.
This means that the $l$-dependence of the loops with
Yukawa couplings can be neglected
to SL accuracy since the on-shell self energy and
vertex diagrams don't produce
an infrared-type contribution in the limit $l \rightarrow 0,
k \rightarrow 0, p_i^2=m_i^2$. Note that to logarithmic
accuracy we can set
$p_i^2=m_s^2$
since all $m_i \leq m_s \sim M$ and since, for now,
we consider only the case of a heavy
photon ($\lambda =M$). The QED type corrections are
included via matching \cite{flmm,habil}
as indicated above. We thus need to
show that the UV-logarithms from the sum of the self energy diagrams
cancel
precisely the UV-logarithms in the sum of the vertex corrections.

The generic one-loop diagrams (modulo the couplings and with common
mass scale $m_s\sim m_t \sim M$), corresponding to the
inner loop insertions depicted in Fig.~(\ref{fig:wi}), are given by:
\begin{eqnarray}
S_{\Sigma} &\equiv&  \int \frac{d^nk}{(2\pi)^n} \frac{ {\rlap/
k}}{(k^2-m_s^2)((k-p_1)^2-m_s^2)}
\label{eq:s1} \\ &=&  \int \frac{d^nk}{(2\pi)^n}
\frac{ ({\rlap/ k}+{\rlap/
p_1})}{(k^2-m_s^2)((k+p_1)^2-m_s^2)}
\label{eq:s2}
\end{eqnarray}
for the self energy insertion and for the vertex diagrams we have for
zero momentum transfer (``soft gauge bosons'', i.e. $l=0$):
\begin{eqnarray}
S^1_{\Lambda_\mu} &\equiv& \int \frac{d^nk}{(2\pi)^n}\frac{ {\rlap/ k}
(2p_1-2k)^\mu}{(k^2-m_s^2)((k-p_1)^2-m_s^2)^2}
\label{eq:v1} \\
S^2_{\Lambda_\mu} &\equiv& \int \frac{d^nk}{(2\pi)^n}\frac{ ({\rlap/
p_1}+{\rlap/ k}) \gamma^\mu
({\rlap/ p_1}+{\rlap/ k})}{(k^2-m_s^2)((k+p_1)^2-m_s^2)^2}
\label{eq:v2}
\end{eqnarray}

Now it is straightforward to obtain from Eq. (\ref{eq:s1}):
\begin{equation}
\frac{\partial S_{\Sigma}}{ {\partial p_1}_\mu}
= - S^1_{\Lambda_\mu} \label{eq:wi1}
\end{equation}
and analogously from Eq. (\ref{eq:s2}):
\begin{equation}
\frac{\partial S_{\Sigma}}{ {\partial p_1}_\mu}
= - S^2_{\Lambda_\mu} \label{eq:wi2}
\end{equation}
Thus, for identical Yukawa couplings, we have to show that
the sum of the couplings to
the various contributing diagrams, multiplying the
{\it same} UV-divergence,
cancel each-other!\par
Indeed this is what happens when one makes on the one hand the sum
of the two loop diagrams with the vertex contributions for the
Yukawa terms (diagrams on the left of Fig.~(\ref{fig:wi}))
and one the other hand the sum of the two loop diagrams with the
self energy insertions (diagrams on the right of Fig.~(\ref{fig:wi})).
This is easy to check in the case of external
right handed top quarks with the exchange of
a photon and a Z-boson.
For each type of (scalar, fermion) virtual contribution
like $(G^\pm, b),~(H^\pm, b),~(\tilde{b},~ \chi^\pm),~(G^0, t),...
~(\tilde{t},~ \chi^0)$
one sees that the same contribution $e^2Q_t^2/c_{\rm w}^2$
due to the sum of the photon and of the Z-boson exchanges
factorizes the Yukawa couplings in the case of
the self energy insertions and in the sum of the two types of 
vertex contributions. In the left handed top quark case,
an analogous result is obtained by considering 
all the left handed diagrams of the process.\par 
As mentioned above, this cancellation is a consequence of the fact that
in spontaneously broken gauge
theories also the Yukawa sector is gauged and that softly
broken supersymmetry preserves the gauge structure.
We can therefore employ the non-Abelian generalization
of the Gribov theorem \cite{gt}
(in the context of the infrared evolution equation method
\cite{flmm,kl}) as indicated in Fig.~(\ref{fig:exp}).
As a result one obtains as a solution of the evolution equation,
for all orders, to SL accuracy, the exponential 
of the universal (DL and
SL) and of the non universal (angular dependent) SL one loop
contributions listed in Section II.
In addition, there appears a contribution arising from the
implementation of the RG effect in the couplings of the
exchanged gauge bosons. This is explicitely discused in ref.\cite{habil}
and the result was already reproduced in eq.(3.6) and (3.9)
of ref.\cite{bmrv}. Using the running expressions, eq.(\ref{aRG}),
of the gauge couplings and expanding to subleading accuracy 
the rather involved 
combinations appearing in the equations of ref.\cite{habil},
\cite{bmrv}, one obtains the additonal $\rm log^3$ term, with the
coefficients $\bar{b}$,
in the following equation.

\begin{eqnarray}
 && d \sigma^{\rm SL}_{e^+_{\alpha} e^-_{\alpha}
\longrightarrow {\overline
f}_{\beta}
f_{\beta}}
= d \sigma^{\rm Born}_{e^+_{\alpha} e^-_{\alpha}
\longrightarrow {\overline f}_{\beta}
f_{\beta}}~~{\rm exp}~\{~
2[\bar{b}_{\alpha}+\bar{b}_{\beta}]
[{1\over3}\log^3({s\over m^2_s})]\nonumber\\
&&
+2(b_{\alpha}+b_{\beta})(2\log{s\over m^2_s}
-\log^2{s\over M^2_V})
+2b^{Yuk}_{\beta}(\log{s\over m^2_s})
+2b^{ang}_{\alpha\beta}(\log{s\over M^2_V})
~\}
\label{eq:fSL}
\end{eqnarray}
\bq
\bar{b}_a={g^4 I_{a}(I_{a}+1)\tilde\beta_0\over64\pi^4}
+{{g^\prime}^4Y^2_{a}\tilde\beta'_0\over256\pi^4}
~~~~~~~~~~~~a=\alpha,\beta
\label{barba}\eq
The coefficients $b$ are extracted from the coefficients $c^{\rm in},~
c^{\rm fin},~c^{\rm Yukawa},~c^{\rm ang}$ in Sect.II:
\bq
b_{a} = \frac{1}{16\pi^2}
\left( g^2 I_{a}(I_{a}+1)+
  {g^\prime}^2\frac{Y^2_{a}}{4}\right)~~~~~~~~~~~~a=\alpha,\beta
\label{ba}\eq
\bq
b^{Yuk}_{\beta} =  - \frac{ g^2}{16 \pi^2} \left(
\frac{1+\delta_{\beta,{\rm R}}}{2}
\frac{{\hat m}
 ^2_f}{M^2_V} + \delta_{\beta,{\rm L}}
 \frac{{\hat m}^2_{f^\prime}}{2 M^2_V} \right)
\label{byuka}\eq

and
\bqa
b^{\rm ang}_{\alpha\beta} &=& -\frac{g^2}{16\pi^2}
\left[
\left( \tan^2
\vartheta_{\rm w}
Y_{e^-_{\alpha}} Y_{
f_\beta}
+ 4 I^3_{e^-_{\alpha}} I^3_{f_\beta} \right)
\log \frac{t}{u} \right. \nonumber
\\ &&
\left. \left. + \frac{\delta_{\alpha, L}
\delta_{\beta,L}}{\tan^2 \vartheta_{\rm w}
Y_{e^-_{\alpha}} Y_{
f_\beta} /4
+ I^3_{e^-_{\alpha}} I^3_{f_\beta}}
\left( \delta_{d,f} \log \frac{-t}{s} -
\delta_{u,f}
\log \frac{-u}{s} \right)  \right] \right\}
\eqa
\noindent
where $I_a$ denotes the total weak isospin of the particle with
chirality $a$, $Y_a$ its weak
hypercharge, ${\hat m}^2_f=$ and 
${\hat m}^2_{f^\prime}=$.
 It should be noted that the one-loop RG corrections do
not exponentiate and are omitted
 in the above expressions. They are, however, completely
determined by the renormalization group
 in softly broken supersymmetric theories such as the MSSM
 and sub-subleading at the higher than one loop order.
The one loop value is the already mentioned RG term Eq.~(\ref{RGder}).

 In Eq. (\ref{eq:fSL}) we use the SUSY-mass scale $m_s$
in the Yukawa and SL-RG terms, but not
in the gauge terms. Using the wrong scale in the DL-type
corrections would unavoidably lead to
wrong SL contributions.

The already existing~\cite{bmrv} explicit results for sfermion
production can be obtained in the straightforward way by using the
corresponding expressions of the various coefficients $b$.

In order to generalize the above results to arbitrary
``light'' SUSY processes
it is convenient to
work in the symmetric basis, i.e. in terms of the symmetry eigenstates.
This is particularly important
for the chargino production that we discussed
at one loop in the previous section.
In the general case
let us denote physical particles (fields) by $f$
and particles (fields) of the unbroken theory by $u$.
Let the connection between
them be denoted by $f=\sum_{u} C^{fu}u$, where the sum
is performed over appropriate
particles (fields) of the unbroken theory.
Note that, in general,
physical particles, having definite masses,  don't belong to
irreducible representations of the symmetry
group of the unbroken theory (for example, the photon and
$Z$ bosons have no definite isospin). On the other hand, particles of
the unbroken theory, belonging to irreducible representations of the
gauge group, have no definite masses.
Then for the amplitude
${\cal M}^{f_1,...f_n}(\{p_k\},\{m_l\};M,\lambda)$ with $n$ physical
particles $f_i$ with momenta $p_i$ and infrared cut-off
$\lambda$, the general case for virtual corrections is given by
\begin{equation}
{\cal M}^{f_1,...f_n}(\{p_k\},\{m_l\};M,\lambda) =  \sum_{u_1,..
.u_n }
\prod_{j=1}^n C^{f_ju_j}
{\cal M}^{u_1,...u_n}(\{p_k\},\{m_l\};M,\lambda) \label{eq:lc}
\end{equation}
In the following we give only the corrections for a
light SUSY mass scale $m_s
\sim M$ and for
a heavy photon ($\lambda = M$) with all $|2p_lp_k| \gg m^2_s, M^2$.
In this case, we can easily work in the symmetric
basis and give the results for these amplitudes. As discussed above and
described in detail in Refs. \cite{flmm,habil,bmrv,bmrv2}, the
soft virtual and real QED corrections must be added by
matching at the weak scale $M$.
It should be mentioned, however, that the Yukawa terms
are independent of the matching terms
and that the Ward identities of the type (\ref{eq:wi1})
and (\ref{eq:wi2}) now apply to the amplitudes
${\cal M}^{u_1,...u_n}(\{p_k\};M)$.

Under these assumptions, we have for general on-shell
matrix elements with $n$-arbitrary external
lines the following resummed SL corrections in the symmetric basis:
\begin{eqnarray}
&& {\cal M}_{\rm SL}^{u_{i_1},...,
u_{i_n}} \left( \{ p_k \}; m_s ; M \right) = \exp \Bigg\{ \sum_{k=1}^n
- \frac{1}{2} \left( b_k(\log^2 \frac{s}{M^2}-2\log\frac{s}{M^2})
 \right)_{i_k \neq
\{W^j,{\widetilde W},B,
{\widetilde B} \}}\nonumber \\
&&
- \frac{1}{2} (b_k\log^2 \frac{s}{M^2})_{i_k =
\{W^j,{\widetilde W},B,
{\widetilde B} \}}
+ \frac{1}{6} (\bar b_k\log^3 \frac{s}{m_s^2})
+b^{PR}_k\log \frac{s}{m_s^2}|_{i_k =
\{W^j,{\widetilde W},B,
{\widetilde B} \}}\nonumber \\
&&
 +\frac{1}{2}b^{\rm Yuk}_{k} \log \frac{s}{m_s^2}
 +b^{\rm ang}_{lk} \log \frac{s}{M^2}
\Bigg\}
{\cal M}^{u_{i_1},...,u_{i^\prime_k},...,u_{i^\prime_l},...,u_{i_n}}_{\rm Born}
(\{p_{k}\})
\label{eq:angr}
\end{eqnarray}
\noindent
with $b_k, \bar{b}_k, ~b^{Yuk}_k$ defined in
eq.(\ref{barba}-\ref{byuka}) and
\bq
b^{PR}_k={g'^2\tilde\beta'_0\over8\pi^2}(\delta_{i_k,B}+
\delta_{i_k,\tilde{B}})+{g^2\tilde\beta_0\over8\pi^2}(\delta_{i_k,W}+
\delta_{i_k,\tilde{W}})
\eq
\bqa
b^{\rm ang}_{lk}&=&\frac{1}{8 \pi^2} \sum^n_{l < k}
\sum_{V_a=B,W^j} \!\!\! {\tilde
I}^{V_a}_{i^\prime_k,i_k}
{\tilde I}^{
{\overline V}_a}_{i^\prime_l,
i_l}  \log \frac{2 p_lp_k}{s}
\eqa
The fields $u$ have a well defined isospin, but for angular
dependent terms involving
CKM mixing effects, one has to include the extended isospin
mixing appropriately in
the corresponding couplings ${\tilde I}^{V_a}_{i^\prime_k,i_k}$
of the symmetric basis.
If some of the sparticles should be heavy, additional
corrections of the form
$\log^2 \frac{m_s^2}{M^2}$ etc. would be important.

The result in Eq. (\ref{eq:angr}), is valid for arbitrary
softly broken supersymmetric extensions
of the SM with the appropriate changes in the $\beta$-functions.
Taking the SUSY-QCD limit
($\frac{g^2}{4 \pi^2} \rightarrow \alpha_s, g^\prime
\rightarrow 0, I_g (I_g+1)
=
I_{\tilde g} (I_{\tilde g}+1) \rightarrow C_A=3,
I_q(I_q+1)=I_{\tilde q} (I_{\tilde q}+1) \rightarrow C_F=4/3,
n_h=0, M=\lambda_g= m_{\tilde g},
C^{\rm yuk}_{i_k}=0$)
of the various terms, Eq. (\ref{eq:angr}) is also valid for the virtual
SUSY-QCD results. It should
be emphasized, however, that in this case the virtual corrections
are not
physical in the sense
that the gluon mass is zero and thus we would need to add the virtual
matching and real contributions before
we could make predictions for collider experiments, while
in the SM soft QED
energy cuts can
define an observable and the heavy gauge boson masses are physical.
In any case, the {\it form} of the operator
exponentiation in color space agrees with
the dimensionally regularized terms in Ref. \cite{cat}
for non-SUSY QCD.\par

The general result in Eq. (\ref{eq:angr}) agrees on the
cross section level with the specific
cross section expressions for fermion
in Eqs. (\ref{eq:fSL}) and similar ones for sfermions and charged
Higgses, as well as
with the one loop results for chargino 
production presented in the previous section.
The appropriate mixing matrices $C^{fu}$ and RG-terms must
be included for this comparison
at one loop. They are sub-subleading at higher orders.\par

We have now at our disposal the sub-leading expressions
for all the considered final states, both at the one-loop
level and completely resummed. Our next goal is that of
comparing the two approximations at variable energy and
verify whether and where they can be considered as
"essentially" (i.e. at the expected one percent experimental
accuracy level) equivalent or, in the same spirit,
"drastically" different. This will be done in the forthcoming
Section 4.\\

\section{Comparison of the two approximations at
subleading \\ order accuracy}

In this Section we evaluate numerically the basic observables (cross section, 
forward-backward
and left-right asymmetries), whose expression is given in Appendix~A, when the 
final state is $t\bar t$, $b\bar b$ and when it is
a pair of charged ``genuinely'' supersymmetric partners, i.e., sfermions, 
charged Higgs and charginos.
For simplicity, we do not report the analysis for production of leptons or light quarks. In these
cases, the previous analytical expressions apply, with the simplification that
the Yukawa terms can be neglected. The numerical results are comparable with the cases 
considered here.
We compute the full effect in the observables in two approximate asymptotic approaches. 
First, we consider the complete set of
one loop Sudakov contributions. As we have explained, these are terms growing 
like $\log^2\frac{s}{M_{\rm W}}$
or $\log\frac{s}{M_{\rm W}}$. Then, we compare this result with the one that is 
obtained by
resumming to all orders at subleading logarithmic accuracy. We remind that this 
corresponds to using an expression
that predicts rigorously all the terms of the form $\alpha^L \log^{2L}(s/M_{\rm 
W}^2)$ and
$\alpha^L \log^{2L-1}(s/M_{\rm W}^2)$.

The deviation on the total cross section, the forward-backward asymmetry
and the left-right asymmetry are defined as
\ba
\Delta\sigma_{tot} &=& \frac{\sigma_{\rm tot}}
{\sigma^{\rm Born}_{\rm tot}} -1 \\
\Delta A_{\rm FB} &=& \frac{\sigma_{\rm FB}}
{\sigma_{\rm tot}} - \frac{\sigma_{\rm FB}^{\rm Born}}
{\sigma^{\rm Born}_{\rm tot}} \\
\Delta A_{\rm LR} &=& \frac{\sigma_{\rm LR}}
{\sigma_{\rm tot}} - \frac{\sigma_{\rm LR}^{\rm Born}}
{\sigma^{\rm Born}_{\rm tot}}
\ea
where $\sigma_{\rm tot}$, $A_{\rm FB}$, $A_{\rm LR}$
are the radiatively corrected observables. If we consider
for instance the total cross section,
in the perturbative one-loop scheme we have
\be
\sigma_{\rm tot} = \sigma^{\rm Born}_{\rm tot}
+ \sigma^{\rm one\ loop}_{\rm tot},
\ee
and in the resummation scheme we must use the expressions that
we have described in the previous Section.

For each final state and for both the one-loop scheme and the resummed scheme we 
consider two
values of the important parameter $\tan\beta$ that we choose to be 
$\tan\beta=10$ and $\tan\beta=40$.
This will allow a discussion of the role of the phenomenologically important 
Yukawa terms and of the validity of the two approaches.

The one-loop Sudakov terms are quadratic and linear
(and the resummed expressions are based on them).
At subleading accuracy we need not specify the scale of the linear logs. The 
scale of the quadratic
ones is conversely important at this level of accuracy. In conclusion, the 
choice of scales can be numerically relevant and we explain how we fixed it.
\begin{enumerate}
\item[] \underline{quadratic logarithms}: these are terms of purely gauge origin 
and their scale
is predicted unambiguously by expanding the full one loop calculation. It is 
$M_W$ or $M_Z$ or the infrared
photon mass regulator $M_\gamma$ according to which of the various diagrams 
originating them is considered.
For the aim of our discussion (mainly the comparison with the resummation 
approach) we have used $M_W$
(called $M_V$ in the Appendices) in all such terms.

\item[] \underline{linear logs of gauge origin}: these are the logarithms that 
combine with the quadratic
ones in order to reconstruct the $2\log-\log^2$ combination. Here, we take 
$M_V=M_W$  for the same previous
reasons.

\item[] \underline{non universal linear logs}: these are single logarithms 
multiplying
non-trivial functions of the scattering angle. Again, these are originated by 
diagrams where the correct
scale can be taken to be $M_V=M_W$.

\item[] \underline{linear RG logs}: here $M_V=M_W$.

\item[] \underline{SUSY logs}: these are single logarithms of Yukawa type 
originated by diagrams with exchange
of SUSY partners. We use a common scale $M_{\rm SUSY}$ fixed at $M_{\rm SUSY} = 
300\ \mbox{GeV}$.
Here, the choice of $M_{\rm SUSY}$ is rather arbitrary, the difference being a 
sub-subleading
term constant with respect to the energy. Our choice is motivated by the recent
investigation~\cite{fulloneloop} where it is shown that this value is a typical 
one for scenarios
where the Sudakov expansion can represent accurately the MSSM effects at 
energies above 1 TeV.
\end{enumerate}

After these preliminary remarks, we turn to the discussion of the numerical 
results and in particular, of their
reliability. With this purpose, we try to sketch a
preliminary qualitative evaluation of the expectable accuracy of 
the resummed expressions.
For simplicity, let us 
consider just the gauge logarithms, that is the combination
$2\log-\log^2$.
In the one loop approximation, the correction is of the form:
\be
\mbox{relative\ effect\ at\ one\ loop} = 1 + \delta_{\rm one\ loop},\quad
\delta_{\rm one\ loop} = c_{\rm 1L}\cdot \alpha(2\log\frac{s}{M_W^2} 
-\log^2\frac{s}{M_W^2}) ,
\ee
where $c_{\rm 1L}$ is a numerical constant. Exponentiating this term would produce, at two loops,
the extra correction $\frac 1 2 \delta_{\rm one\ loop}^2$.
Since the resummation is able to identify correctly only the leading and subleading
 logs at all orders, but not  the sub-subleading terms,  we would have for the complete
two-loop correction an expression of the kind:
\be
\mbox{ real\ relative\ resummed\ effect} = 1 + \delta_{\rm one\ loop} + \frac 1 
2  \delta^2_{\rm one\ loop}
\pm c_{\rm R}^2 \alpha^2 \log^2\frac{s}{M_W^2} + \cdots
\ee
where we have used in the theoretical error the smallest scale that we have. The 
dominant theoretical uncertainty of the
resummation procedure can then be estimated at 3 TeV as
\be
\pm c_{\rm R}^2\alpha^2\log^2\frac{s}{M_W^2} = \pm \frac{c_{\rm R}^2}{c_{\rm 
1L}^2}
\delta_{\rm one\ loop}^2\cdot \frac{\log^2\frac{s}{M_W^2}}
{(2\log\frac{s}{M_W^2} -\log^2\frac{s}{M_W^2})^2} \simeq \pm \frac 1 {28}\ 
\frac{c_{\rm R}^2}{c_{\rm 1L}^2}\
 \delta^2_{\rm one\ loop}
\ee
Hence, the dominant theoretical uncertainty of the resummation expressions (at 
subleading accuracy)
would be below 1\% if the one loop effect were below 50\%, 25\% or  10 \% for $c_{\rm 
R}/c_{\rm 1L} = 1.1, 2.1, 5.3$
respectively. Since potentially large contributions to $c_{\rm R}$ can in principle
appear due to terms that are already present
at one loop, (e.g. relatively large non logarithmic - constant - terms like those 
found in \cite{bmrv2} for charged Higgs production)
and since the above analysis
is admittedly naive, our conservative attitude would be for the moment that of
not considering the
resummation expressions as the final word for a high accuracy prediction when
the one loop effect is beyond, say, 10\%, in which case we feel that a (tough!)
partial two-loop calculation would be highly desirable.\\

In the remaining part of this Section, for each final states, we shall plot the 
effects in the two
asymptotic approximate approaches as function of the energy for two values of 
$\tan\beta=10,40$. We also collect
in Tab.~(\ref{finaltable}) the results for $\tan\beta=10$ at 1 and 3 TeV. The list of the considered
case is now following.

\bigskip

\noindent\underline{final top and bottom}

\noindent
The effects are shown in Figs.~(\ref{fig:top},\ref{fig:bottom}). At 1 TeV, the 
one-loop effects in
$\sigma_t$ and $\sigma_b$ are still, essentially, within the assumed limit of 
10\%, and their differences with respect to the resummed expansions are of approximately one 
percent. Similar conclusions apply for the two considered asymmetries. To the subleading logarithmic
accuracy, a one-loop calculation seems therefore sufficient for all the considered observables at that
energy, i.e. it does not ``practically'' change after (the corresponding) resummation. At 3
TeV, the one-loop effect on $\sigma_t$ is
definitely beyond the ten percent value, the difference with respect to the resummed
expressions reaches a ten percent size (note the reduction of the size of the resummed effect). Also
the two asymmetries show visible differences between the two approaches (and $A_{LR}$ is particularly
large at one-loop) at that energy, as one sees from the Figure. The situation appears
definitely better for bottom production, where both the one-loop effects and the differences
with respect to the resummed expressions remain reasonably small (also, it is conceivable that in this
case the available experimental accuracy is worse than in the top case). \\

\noindent\underline{final sfermions and charged Higgs}

\noindent
The effects are shown in Figs.~(\ref{fig:stopleft}-\ref{fig:higgs}). At 1 TeV 
all the observables
are "under control" with quite small correction from the resummation procedure. 
At 3 TeV, there are
problematic large corrections to
$\sigma_{{\tilde t}_L}$, $A_{FB,{{\tilde t}_L}}$, $\sigma_{{\tilde t}_R}$, 
$\sigma_{{\tilde b}_L}$,
$A_{FB,{{\tilde b}_L}}$, $\sigma_{H}$, $A_{FB,H}$. In particular, for the cross 
sections
$\sigma_{{\tilde t}_L}$, $\sigma_{{\tilde t}_R}$, $\sigma_{{\tilde b}_L}$,
and $\sigma_{H}$ the effects are around 20 \% and raise, in our opinion, serious 
doubts about the reliability of the
resummation at this level of accuracy.

\noindent\underline{final charginos}

\noindent
If the final state is a pair of chargino and antichargino, we can in principle 
consider three separate
cases according to which charginos are produced. In other words we consider the 
process
\be
e^+e^- \to \chi_a^+ \chi_b^-
\ee
in the three cases $(a,b) = (1,1), (1,2), (2,2)$. In this case the observables 
depend
on the mixing between the Higgsino and the Wino component in the mass 
eigenstates charginos.
The chargino mixing matrix depend on the MSSM standard parameters $M_2$, $\mu$ 
and $\tan\beta$.
We have chosen to evaluate the observables at $M_2 = 200$ GeV and $\mu = 300$ 
GeV as a representative
light scenario. In a forthcoming paper, we shall discuss in full details the 
dependence on the mixing
focusing on the constraints that virtual correction impose on the determination 
of the chargino
system parameters.
The effects are shown in Figs.~(\ref{fig:obs11}-\ref{fig:obs22}).
In the phenomenological analysis, it is sometimes reasonable to assume that only 
the lightest chargino
has been produced asking for what information can be gained in this case. 
However, especially in a light
SUSY scenario, one can also consider a more favorable situation where both 
charginos can be
produced. Then, one can study the inclusive observables like, for instance,
\be
\sigma_{\rm inc} = \sigma_{11}+2\sigma_{12}+\sigma_{22}
\label{inclusive}
\ee
for which the Sudakov expansion is completely independent on the mixing and 
there is no dependence
on both $M_2$ and $\mu$. The effects in this representative example are shown in 
Fig.~(\ref{fig:inclusive}).
Looking at Tab.~(\ref{finaltable}) we see that at 1 TeV the asymmetries have corrections
below 10\% and the cross section is just above that value. The effect of resummation 
is in the range of 1-2 percents for $\sigma$ and $A_{FB}$ and one order of magnitude smaller
in $A_{LR}$.
At 3 TeV, on the contrary, both the cross section and the forward-backward asymmetry are large
and the resummation introduces large shifts, particularly in the cross-section (+10\%).
A detailed analysis reveals that the large effects are mainly due to the Wino 
component where, at one loop,
there is not single logarithm compensating the leading squared logarithm.

\section{Conclusions}

In this paper we have considered all the processes of production of  pairs 
of charged particles with spin zero and one-half from electron-positron 
annihilation at TeV energies in the framework of a given (MSSM) 
supersymmetric model. 
Assuming a (relatively) light SUSY scenario, with typical
values for the relevant masses below a few hundred GeV, we have concentrated our 
attention on the
asymptotic Sudakov logarithmic expansion of the electroweak component of the 
invariant scattering 
amplitude. We have computed at one-loop all the leading Double logarithms and 
the subleading
Linear logarithms of the expansions, also including the linear logarithms of RG 
origin. To
subleading logarithmic accuracy, we have also computed, for the same processes, 
all the resummed
exponentiated expressions. To our knowledge, this is the first complete 
calculation of this kind,
and we do not have at disposal  calculations of different authors with which to 
compare our
expressions. We have, though, performed an internal self-consistency check of 
our results for the
chargino case, and we hope that there are no mistakes in our several formulae.\\
Given the two approximate asymptotic expressions, we have verified that, for 
energies in the one
TeV range (final goal of a future LC), there are practically never visible (in 
our working assumptions, beyond a ~one percent
level) differences between the two results for a set of realistic experimental 
observables. On the
 contrary, rather strong discrepancies appear almost systematically as soon as 
one approaches the 
 few (2,3) TeV regime (goal of a future CLIC).\\
 A first conclusion is therefore that, in the one TeV range, an asymptotic 
Sudakov expansion
 at one loop has a subleading logarithmic accuracy that does not require extra 
resummation, for
 the considered processes. The same conclusion cannot evidently
be drawn in the higher 
energy considered
 regime.\\
 We stress at this point two important facts. The first one
is that the validity to subleading
 accuracy of the one-loop expansion does not necessarily
mean that it is completely accurate when one moves beyond that level of
accuracy. Extra terms, in
 particular energy-independent ones, might well be relevant at the one percent 
level in the
 expansion. The second one is that the validity of the expansion itself,
i.e. the fact that it gives indeed an adequate description of the real
(complete) effect even after inclusion of
 possible extra e.g. constant terms, remains to be demonstrated.\\
 It is rather simple to realize that the two points are related, and one 
possible solution would
 be represented by the preparation of a full one-loop program. This would allow 
to check the
 validity of a logarithmic expansion, at the same time allowing to determine by 
a proper fit to
 the complete result the value of a possible extra constant term.\\
 In the special case of charged Higgs production, this task has been actually 
carried through~\cite{bmrv2}. One can see therefore that for 
that process an asymptotic 
Sudakov expansion at
 one loop, with an additional constant term to be realistically estimated, is 
suitable at one TeV.
 The benefit of this conclusion is represented by the fact that, in such an 
improved expansion, the
 coefficients of the different asymptotic terms depend each one
on special reduced subsets of the supersymmetric
 parameters. This would allow, via identification of the various terms, 
stringent and simple tests of the model, of which one must 
assume the previous 
 experimental confirmation. In fact, a declared goal of future colliders
will be also that of
 performing precision tests of the (assumedly discovered ) supersymmetric 
model.\\
 In the case of charged Higgs production, the coefficient of the linear Yukawa 
logarithm was \underline{only dependent on $\tan\beta$}. 
For the latter, particularly if its value were large (beyond, say, 
ten), the precise experimental
 determination is not at the moment completely clean, and we proposed in 
previous papers \cite{bmrv,brv3,fulloneloop} to measure $\tan\beta$
 from the slope of the final pair cross section. In the case of e.g. chargino 
production, other SUSY parameters
 e.g. of mixing type
 would enter the coefficients of the subleading logarithms, and 
different ones will appear in next-to subleading terms, so that
another independent
relevant test of the model would be available. To perform the test in a
rigorous way would
 require the preparation of a complete one-loop program, analogous to the one 
completed for Higgs production. From
 what shown in this paper, we would hope that, 
in a relatively light SUSY scenario, a logarithmic one-loop expansion (with a
possible addition of an extra constant term) were able to provide an adequate
description of chargino production in the
one TeV range, and no need of hard two-loop 
calculations were advocated.
We are already working along that direction.

\newpage

\appendix

\section{Observables for production of fermionic and of scalar pairs}

\subsection{Fermionic pairs}

For any type of fermionic pair (leptons, quarks, charginos,
neutralinos) we can write the complete amplitude as:
\bq
A(e^+e^-\to i j)\equiv\sum_{ab} A^{ab}
(\gamma^\mu P_a)^{ee}(\gamma_\mu P_b)^{ij}
\label{Aab}\eq
\noindent
with $a,b=L$ or $R$ and
\bq
(\gamma^\mu P_a)^{ee}=\bar{v}(e^+)\ \gamma^{\mu}P_a\ u(e^-)
~~~~~~~
(\gamma_\mu P_b)^{ij}=\bar{u}(i)\ \gamma_{\mu}P_a \ v(j)
\eq

The \underline{unpolarized angular distribution} is given by
\bqa
{d\sigma\over d\cos\vartheta}&=&{1\over32\pi s}~\left[u^2
\left(|A^{RR}|^2+|A^{LL}|^2\right)+t^2\left(|A^{LR}|^2+|A^{RL}|^2\right)\right]
\eqa
\noindent
 and the
\underline{Left-Right polarization asymmetry}
is obtained as

 \bq
A_{LR}(s, \vartheta) =
[{d\sigma^{LR}\over d\cos\vartheta}]/[{d\sigma\over d\cos\vartheta}]
 \label{ALR}\eq
with
\bqa
{d\sigma^{LR}\over d\cos\vartheta}&=&{1\over32\pi s}~\left[u^2
(|A^{LL}|^2-|A^{RR}|^2)+t^2(|A^{LR}|^2-|A^{RL}|^2)\right]
\eqa

The \underline{Forward-Backward asymmetry} is

\bq
{A}_{FB}= {(\int_F-\int_B){d\sigma\over d\cos\vartheta}\over
(\int_F+\int_B){d\sigma\over d\cos\vartheta}}
\label{afb}\eq
where $\int_F = \int_0^1 d\cos\vartheta$ and $\int_B = \int_{-1}^0 
d\cos\vartheta$.

\subsection{Scalar pairs}

Following the notations of ref.\cite{bmrv} we write the amplitude as:
\bq
A \equiv {2e^2\over s}~
\bar v(e^+)\ \gamma^{\mu} p_{\mu}~ (a_LP_L+a_RP_R)~u(e^-)
\eq
The \underline{unpolarized cross section} is
\bq
{d\sigma\over
d\cos\vartheta}=N_{col}{\pi\alpha^2\over8s}~\sin^2\vartheta~
(|a_{L}|^2+|a_{R}|^2)
\eq
\noindent
whereas
the \underline{Left-Right polarization asymmetry} is obtained
from eq.(\ref{ALR}) with
\bq
{d\sigma^{LR}\over
d\cos\vartheta}=N_{col}{\pi\alpha^2\over8s}~\sin^2\vartheta~
(|a_{L}|^2-|a_{R}|^2)
\eq
\noindent
and the \underline{Forward-Backward asymmetry}
is still given by (\ref{afb}).

\newpage

\section{Amplitudes at one loop }

\subsection{Leptons or quarks}
The Born amplitudes for a final fermion pair $f\bar{f}$,
with the notations of Appendix A are
\bq
A^{\rm Born}_{LL}=
{e^2\over 4s^2_{\rm w}c^2_{\rm w} s}[(2s^2_{\rm w}-1)(2I^3_f)-2s^2_{\rm w} Q_f]
~~~~~~~
A^{\rm Born}_{LR}=
-~{e^2\over 2c^2_{\rm w} s}[Q_f]
\eq
\bq
A^{\rm Born}_{RL}=
{e^2\over c^2_{\rm w} s}[I^3_f-Q_f]
~~~~~~
A^{\rm Born}_{RR}=
-~{e^2\over c^2_{\rm w} s}[Q_f]
\eq

The one loop terms, factorizing these Born terms
are given in Section II.

\subsection{Sleptons, squarks or charged Higgs bosons}

With the notations of Appendix A and ref.\cite{bmrv}, the Born
amplitudes for a final sfermion pair $\tilde f\bar{\tilde f}$ are

\bq
a^{\rm Born}_L=-Q_f+{(I^3_f-s^2_{\rm w}Q_f)g_{eL}\over2s^2_{\rm w}c^2_{\rm w}}=
-~{s^2_{\rm w} Q_f+(1-2s^2_{\rm w})I^3_f\over2s^2_{\rm w}c^2_{\rm w}}
\eq
\bq
a^{\rm Born}_R=-Q_f+{(I^3_f-s^2_{\rm w}Q_f)g_{eR}\over2s^2_{\rm w}c^2_{\rm w}}=
{I^3_f-Q_f\over c^2_{\rm w}}
\eq

This applies to $H^+H^-$ using $Q_f=-1$, $I^3_f=-{1\over2}$ and
$Y_f=-1$, i.e.
\bq
a^{\rm Born}_L={1\over4s^2_{\rm w}c^2_{\rm w}}
~~~~~
a^{\rm Born}_R={1\over2c^2_{\rm w}}
\eq

The one loop terms, factorizing these Born terms
are given in Section II.

\subsection{Charginos $\chi^{+}_i\chi^-_j$}

At one loop, following the notations of Appendix A,
the amplitudes $A^{ab}_{ij}$,
where $ab$ refer to $LL,LR,RL,RR$, originate from $s,u,t$ channel
contributions:

\bq
A^{ab}_{ij}\equiv{e^2\over s}S^{ab}_{ij}+{e^2\over u}U^{ab}_{ij}
+{e^2\over t}T^{ab}_{ij}
\eq
\noindent
with
\bq
S^{ab}_{ij}= S^{ab,~Born}_{ij}
+S^{ab,~Born}_{ij} c^{in}_{a}
+\sum_k S^{ab,~Born}_{ik} c^{fin}_{kj}+S^{ab,~ang}+S^{ab,~RG}_{ij}
\eq

\bq
U^{LL}_{ij}= U^{LL,~Born}_{ij}
+U^{LL,~Born}_{ij} c^{in}_{L}+\sum_k
U^{LL,~Born}_{ik} c^{fin}_{kj}+U^{LL,~ang}_{ij}
\eq

\bq
T^{LR}_{ij}=T^{LR,~ang}_{ij}
\eq
\noindent
in which we have specified the contributions of the Born terms (photon
and $Z$ exchange in the $s$ channel, sneutrino exchange in the $u$
channel),
the universal corrections from initial and final
lines, the angular dependent corrections and the RG
corrections. To make them
explicit, it is convenient to separate the Higgsino and the Wino parts:

\bq
A^{ab}_{ij}=A^{ab,\rm Hig}_{ij}+A^{ab,\rm Win}_{ij}
\label{ahigwin}\eq
At Born level, the only non vanishing terms are:
\bq
\ S^{LL~ \rm,Hig~ Born}_{ij}=-~{1\over4s^2_{\rm w}c^2_{\rm 
w}}Z^{+*}_{2i}Z^+_{2j}
~~~~~~~
\ S^{LL~ \rm,Win~ Born}_{ij}=-~{1\over2s^2_{\rm w}}Z^{+*}_{1i}Z^+_{1j}\eq
\bq
\ U^{LL~ \rm,Win~ Born}_{ij}
=-~{1\over2s^2_{\rm w}}Z^{+*}_{1i}Z^+_{1j}
\eq
\bq
\ S^{LR~ \rm,Hig~ Born}_{ij}=-~{1\over4s^2_{\rm w}c^2_{\rm 
w}}Z^{-}_{2i}Z^{-*}_{2j}
~~~~~~~~~~~
\ S^{LR~ \rm,Win~ Born}_{ij}=-~{1\over2s^2_{\rm w}}Z^{-}_{1i}Z^{-*}_{1j}
\eq
\bq
\ S^{RL~ \rm,Hig~ Born}_{ij}=-~{1\over2c^2_{\rm w}}Z^{+*}_{2i}Z^+_{2j}
\eq
\bq
\ S^{RR~ \rm,Hig~ Born}_{ij}=-~{1\over2c^2_{\rm w}}Z^{-}_{2i}Z^{-*}_{2j}
\eq
\noindent
such that, at one loop, one can write:
\bqa
A^{ab,\rm Hig}_{ij}&=&{e^2\over s}S^{ab~ \rm,Hig~ Born}_{ij}
\{~1+(b^{in}_a+b^{fin,Hig})(2\log{s\over M^2}-\log^2{s\over M^2_V})
\nonumber\\
&&
+b^{Yuk}_b(\log{s\over M^2})+b^{ang,Hig}_{ab}(\log{s\over M^2_V})~\}
+{e^2\over s}S^{ab,\rm Hig,RG}_{ij}
\eqa
\bqa
A^{ab,\rm Win}_{ij}&=&[{e^2\over s}S^{ab~ \rm,Win~ Born}_{ij}+
{e^2\over u}U^{ab~ \rm,Win~ Born}_{ij}]\{~1
+(b^{in}_a)(2\log{s\over M^2}-\log^2{s\over M^2_V})\nonumber\\
&&
+b^{fin,Win}(-\log^2{s\over M^2_{\rm w}})+
b^{ang,Win}_{ab}(\log{s\over M^2})~\}
\eqa
\noindent
with the initial and final "gauge" coefficients
\bq
b^{in}_L= b^{fin,~Hig}=\frac{\alpha}{16\pi s^2_{\rm w}c^2_{\rm w}}( 1+2c^2_{\rm 
w})
~~~~~~~
b^{in}_R=\frac{\alpha}{4\pi c^2_{\rm w}}\eq
\bq
b^{fin,~Win}={\alpha\over2\pi s^2_{\rm w}}
\label{bfinwin}\eq
\noindent
the final Yukawa corrections
\bq
b^{Yuk}_L= -~\frac{3\alpha}{8\pi s^2_{\rm w}}[{m^2_t\over
M^2_{\rm w}}(1+\cot^2\beta)]
~~~~~~~~~
b^{Yuk}_R= -~\frac{3\alpha}{8\pi s^2_{\rm w}}[{m^2_b\over
M^2_{\rm w}}(1+\tan^2\beta)]
\eq
\noindent
the angular dependent terms
\bqa
b^{ang,~Hig}_{LL}&=&b^{ang,~Hig}_{LR}=
-~{\alpha\over4\pi s^2_{\rm w}c^2_{\rm w}}\log({u\over t})
-{\alpha c^2_{\rm w}\over\pi  s^2_{\rm w}}\log({-u\over s})
\eqa
\bqa
b^{ang,~Hig}_{RL}&=&b^{ang,~Hig}_{RR}=
-~{\alpha\over2\pi c^2_{\rm w}}\log({u\over t})
\eqa
\bqa
b^{ang,~Win}_{LL}&=&b^{ang,~Win}_{LR}=
-~{\alpha\over 2\pi s^2_W}~\{~\log({u\over t})
+(1-{u\over t})\log({-u\over s})~\}
\eqa
\noindent
and the RG terms (from the Higgsino components)
\bqa
S^{LL,~RG}=-~{\alpha\over4\pi}[Z^{+*}_{2i}Z^{+}_{2j}]
[({1\over s^4_{\rm w}c^4_{\rm w}})
({3-6s^2_{\rm w}+8s^4_{\rm w}\over6}N_{\rm fam}-{5-10s^2_{\rm w}+4s^4_{\rm 
w}\over4})]
[\log \frac{s}{M^2}]
\nonumber\\
&&\eqa
\bqa
S^{LR,~RG}&=&-~{\alpha\over4\pi}[Z^{-}_{2i}Z^{-*}_{2j}]({1\over s^4_{\rm 
w}c^4_{\rm w}})
({3-6s^2_{\rm w}+8s^4_{\rm w}\over6}N_{\rm fam}-{5-10s^2_{\rm w}+4s^4_{\rm 
w}\over4})]
[\log \frac{s}{M^2}]
\nonumber\\
&&\eqa
\bqa
S^{RL,~RG}&=&-~{\alpha\over4\pi}[Z^{+*}_{2i}Z^{+}_{2j}]
({1\over c^4_{\rm w}})({5\over3}N_{\rm fam}+{1\over2})[\log \frac{s}{M^2}]
\eqa
\bqa
S^{RR,~RG}&=&-~{\alpha\over4\pi}[Z^{-}_{2i}Z^{-*}_{2j}]({1\over c^4_{\rm w}})
({5\over3}N_{\rm fam}+{1\over2})[\log \frac{s}{M^2}]
\eqa

 It is interesting to check the expression of the "gauge"
Wino contribution eq.(\ref{bfinwin}), by making the sum of
the $\chi^{\pm}$ splitting function term
and of the Wino
Parameter Renormalization term, ref.\cite{brv4}:

\bqa
&&2c^{split}_{kj}=
{\alpha\over2\pi s^2_{\rm w}}\{[-\log^2{s\over M^2_{\rm w}}~]
-~(N_{\rm fam}-~{5\over 2})[\log{s\over M^2}]\}
[Z^{+*}_{1k}Z^{+}_{1j}P_L+Z^{-}_{1k}Z^{-*}_{1j}P_R]\eqa

\bqa
&&2c^{PR}_{kj}=[~{\alpha\over 2\pi s^2_{\rm w}}]
~[(N_{\rm fam}-{5\over2})][\log{s\over M^2}]
[Z^{+*}_{1k}Z^{+}_{1j}P_L+Z^{-}_{1k}Z^{-*}_{1j}P_R]
\eqa
\noindent
and observing that all single log cancel and that the total Wino part
is a pure quadratic log, exactly like in $W^+W^-$ production.


\newpage

\section{Resummed Amplitudes}

The case of ordinary fermions has already been given in Section
III, eq.(\ref{eq:fSL}).
The extension to sfermions and Higgses is straightforward
using the corresponding expressions for the various coefficients
$b$. We make now explicit the case of Charginos which is more delicate.

 In correspondence
with eq.(\ref{ahigwin}), following eq.(\ref{eq:angr}), the resummed
amplitude called $B^{ab}$, with $ab=LL,LR,RL,RR$, is written:

\bq
B^{ab}_{ij}=B^{ab,\rm Hig}_{ij}+B^{ab,\rm Win}_{ij}
\eq
\bqa
B^{ab,\rm Hig}_{ij}&=&{e^2\over s}S^{ab~ \rm,Hig~ Born}_{ij}
~{\rm exp}~\{~[\bar{b}^{in}_{a}+\bar{b}^{fin,~Hig}]
[{1\over3}\log^3({s\over M^2})]\nonumber\\
&&
+(b^{in}_a+b^{fin,Hig})(2\log{s\over M^2}-\log^2{s\over M^2_V})
+b^{Yuk}_b(\log{s\over M^2})+b^{ang,Hig}_{ab}(\log{s\over M^2_V})~\}
\nonumber\\
&&+B^{ab,~Hig~RG}_{ij}\eqa
\bqa
B^{ab,\rm Win}_{ij}&=&[{e^2\over s}S^{ab~ \rm,Win~ Born}_{ij}+
{e^2\over u}U^{ab~ \rm,Win~ Born}_{ij}]~{\rm exp}~\{~
[\bar{b}^{in}_{a}+\bar{b}^{fin,~Win}]
[{1\over3}\log^3({s\over M^2})]\nonumber\\
&&
(b^{in}_a)(2\log{s\over M^2}-\log^2{s\over M^2_V})
+b^{fin,Win}(-\log^2{s\over M^2_V})\nonumber\\
&&+
b^{Win~PR} [\log({s\over M^2})]+
b^{ang,Win}_{ab}(\log{s\over M^2_V})~\}\nonumber\\
&&+B^{ab,~Win,~RG}_{ij}
\eqa
\noindent
with the new quantities (not defined in the one loop expression):
\bq
\bar{b}^{in}_L= \bar{b}^{fin,~Hig}
=\frac{3\alpha^2\tilde\beta_0}{16\pi^2 s^4_{\rm w}}+
\frac{\alpha^2\tilde\beta'_0}{16\pi^2 c^4_{\rm w}}
~~~~~~~~
\bar{b}^{in}_R=
\frac{\alpha^2\tilde\beta'_0}{4\pi^2 c^4_{\rm w}}
\eq

\bq
\bar{b}^{fin,~Win}=
{\alpha^2\tilde\beta_0\over2\pi^2 s^4_{\rm w}}
~~~~~~~~b^{Win~PR}={\alpha\tilde\beta_0\over\pi s^2_{\rm w}}
\eq

\bqa
B^{LL,~Hig,~RG}&=&-\{~{e^2(s)\over 4ss^2_{\rm w}(s)c^2_{\rm w}(s)}
-[{e^2\over 4ss^2_{\rm w}c^2_{\rm w}}]_{\rm Born}~\}
[Z^{+*}_{2i}Z^{+}_{2j}]\eqa

\bqa
B^{LR,~Hig,~RG}&=&-\{~{e^2(s)\over 4ss^2_{\rm w}(s)c^2_{\rm w}(s)}
-[{e^2\over 4ss^2_{\rm w}c^2_{\rm w}}]_{\rm Born}~\}
[Z^{-}_{2i}Z^{-*}_{2j}]\eqa

\bqa
B^{RL,~Hig,~RG}&=&-\{~{e^2(s)\over 2sc^2_{\rm w}(s)}
-[{e^2\over 2sc^2_{\rm w}}]_{\rm Born}~\}
[Z^{+*}_{2i}Z^{+}_{2j}]\eqa
\bqa
B^{RR,~Hig,~RG}&=&-\{~{e^2(s)\over 2sc^2_{\rm w}(s)}
-[{e^2\over 2sc^2_{\rm w}}]_{\rm Born}~\}
[Z^{-}_{2i}Z^{-*}_{2j}]\eqa

\bqa
B^{LL,~Win,~RG}&=&-\{~{e^2(s)\over2 ss^2_{\rm w}(s)}
+{e^2(s)\over2 us^2_{\rm w}(s)}-[{e^2\over2 ss^2_{\rm w}}
+{e^2\over2 us^2_{\rm w}}]_{\rm Born}~\}[Z^{+*}_{1i}Z^+_{1j}]
\eqa
\bqa
B^{LR,~Win,~RG}&=&-\{~{e^2(s)\over2 ss^2_{\rm w}(s)}
-[{e^2\over2 ss^2_{\rm w}}
]_{\rm Born}~\}[Z^{-}_{1i}Z^{-*}_{1j}]
\eqa

\bqa
B^{RL,~Win,~RG}&=&B^{RL,~Win,~RG}=0
\eqa
\noindent
where the running expressions depending of $s$ are defined
in eq.(\ref{aRG}).\par
One can check that with the expansion
${\rm exp(X)}=1+{\rm X}$ one recovers
the one loop contribution apart from a {two loop $\alpha^2\log^3$ term
and a reshuffling of the Wino PR terms.

\newpage

\begin{table}[floatfix]
\begin{tabular}{|l|cc|c|cc|c|}
\hline
Observable & 1loop (1 TeV) & res. (1 TeV) & $\Delta$ & 1loop (3 TeV) & res. (3 
TeV) & $\Delta$ \\
\hline
$\sigma_t$                 & {\bf -10}    & -8.8   &   1.2   & {\bf -26}   &  
-21   &   5      \\
$A_{FB,t}$                 & -3.5   & -3.1   & 0.4   & -6.4  & -4.5  &    1.9   
\\
$A_{LR,t}$                 & -4.2   & -3.8   & 0.4   & {\bf -12}   & -8.9  &   
3.1    \\
$\sigma_b$                 & {\bf 9.6}    &  11     &   1.4   & 1.3   & 4.0   &  
  2.7   \\
$A_{FB,b}$                 & 4.8    & 5.1    & 0.3   & 7.1   & 6.7   & -0.4   \\
$A_{LR,b}$                 & 1.5    & 2.0    & 0.5   & -0.54 & 0.18  & 0.72   \\
\hline
$\sigma_{{\tilde t}_L}$    & -4.7   & -4.0   & 0.7   & {\bf -21}   &  -17   &   
4      \\
$A_{FB,{{\tilde t}_L}}$    & -7.0   & -6.5   & 0.5   & {\bf -12}   &  -9.2  &   
2.8    \\
$A_{LR,{{\tilde t}_L}}$    & -0.086 & -0.083 & 0.003 & -0.47 & -0.42 & 0.05   \\
$\sigma_{{\tilde t}_R}$    & -1.9   & -1.4   & 0.5   &  {\bf -9.9}  & -7.6  &   
2.3    \\
$A_{FB,{{\tilde t}_R}}$    & -1.6   & -1.5   & 0.1   & -2.6  & -2.0  &  0.6   \\
$A_{LR,{{\tilde t}_R}}$    & -1.4   & -1.2   & 0.2   & -3.8  & -2.9  & 0.9    \\
$\sigma_{{\tilde b}_L}$    & -4.1   & -3.4   & 0.7   & {\bf -20}   &  -17   &   
3      \\
$A_{FB,{{\tilde b}_L}}$    & 7.2    & 6.9    & -0.3  & {\bf 13}    &  9.8   &   
-3.2   \\
$A_{LR,{{\tilde b}_L}}$    & -0.071 & -0.071 & 0     & -0.58 & -0.55 & 0.03   \\
$\sigma_{{\tilde b}_R}$    & 4.5    & 4.6    & 0.1   & 2.7   & 3.2   & 0.5    \\
$A_{FB,{{\tilde b}_R}}$    & 0.77   & 0.74   & -0.03 & 1.1   & 1.0   & -0.1   \\
$A_{LR,{{\tilde b}_R}}$    & -1.3   & -1.2   & 0.1   & -3.3  & -2.9  & 0.4    \\
$\sigma_{H}$               & -2.8   & -2.2   & 0.6   & {\bf  -17}  &   -14  &   
3      \\
$A_{FB,H}$                 & 5.9    & 5.6    & -0.3  & {\bf 10}    & 7.7   &   
-2.3   \\
$A_{LR,H}$                 & -0.53  & -0.48  & 0.05  & -2.4  & -2.0  & 0.4    \\
\hline
$\sigma_{\rm charginos}$   & {\bf -13}    & {\bf -11}    & 2     & {\bf -40}   & {\bf -30}   &  10    \\ 
$A_{FB,\rm charginos}$     & -6.3   & -5.4   & 0.9   & {\bf -11}   & -6.7  &  4.3   \\
$A_{LR,\rm charginos}$     & -1.0   & -0.79  & 0.21  & -3.8  & -2.2  &  1.6   \\
\hline
\end{tabular}
\caption{Summary  table for the effect at $\tan\beta=10$. The numbers in boldface 
are effects larger than about 10\%.}
\label{finaltable}
\end{table}

\newpage

\begin{figure}[thb]
\centering
\epsfig{file=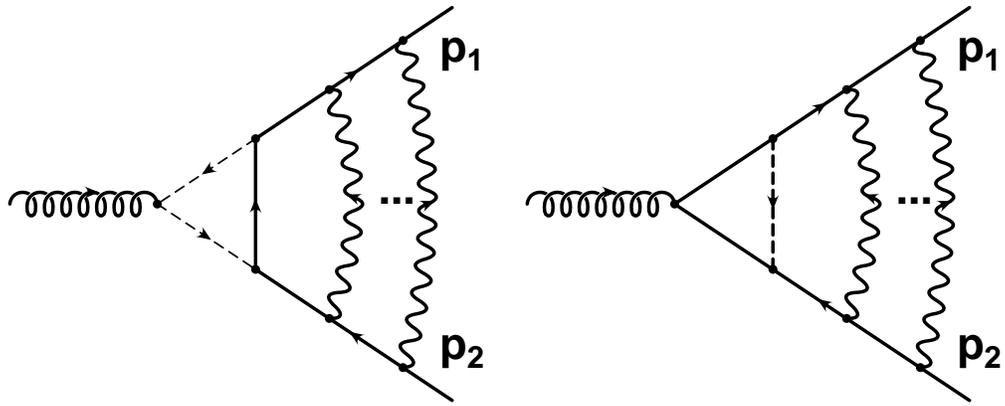,width=0.8\textwidth}
\vspace{1cm}
\caption{Higher order corrections to the original on-shell vertex diagrams with
Yukawa couplings. The non-Abelian generalization of the Gribov theorem can be
applied as is shown in the text. The figure is only schematic since in principle 
the gauge
bosons couple to all external legs.} \label{fig:exp}
\end{figure}

\begin{figure}
\centering
\epsfig{file=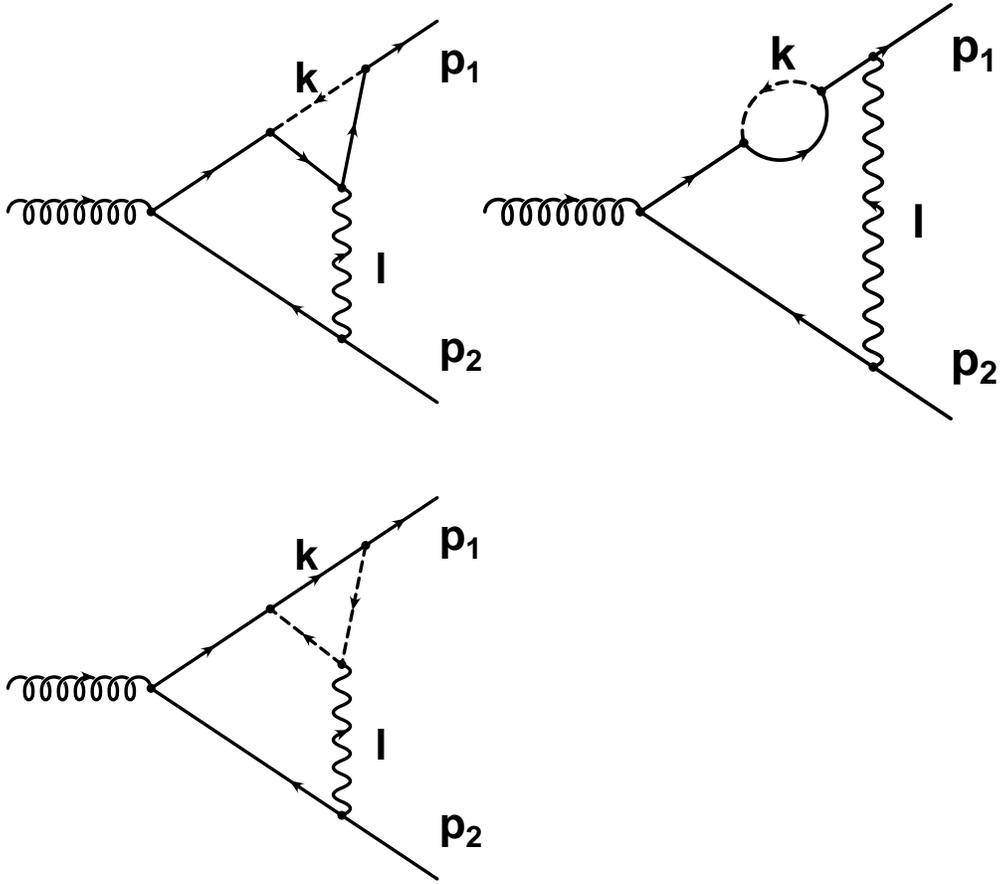,width=0.8\textwidth}
\vspace{1cm}
\caption{Two loop Feynman diagrams involving Yukawa couplings. In the text it is 
shown that
the sum of all such contributions with self energy subloops are canceled by the 
sum
of the corresponding vertex diagrams to SL accuracy.} \label{fig:wi}
\end{figure}

\begin{figure}[tb]
\begin{center}
\leavevmode
\epsfig{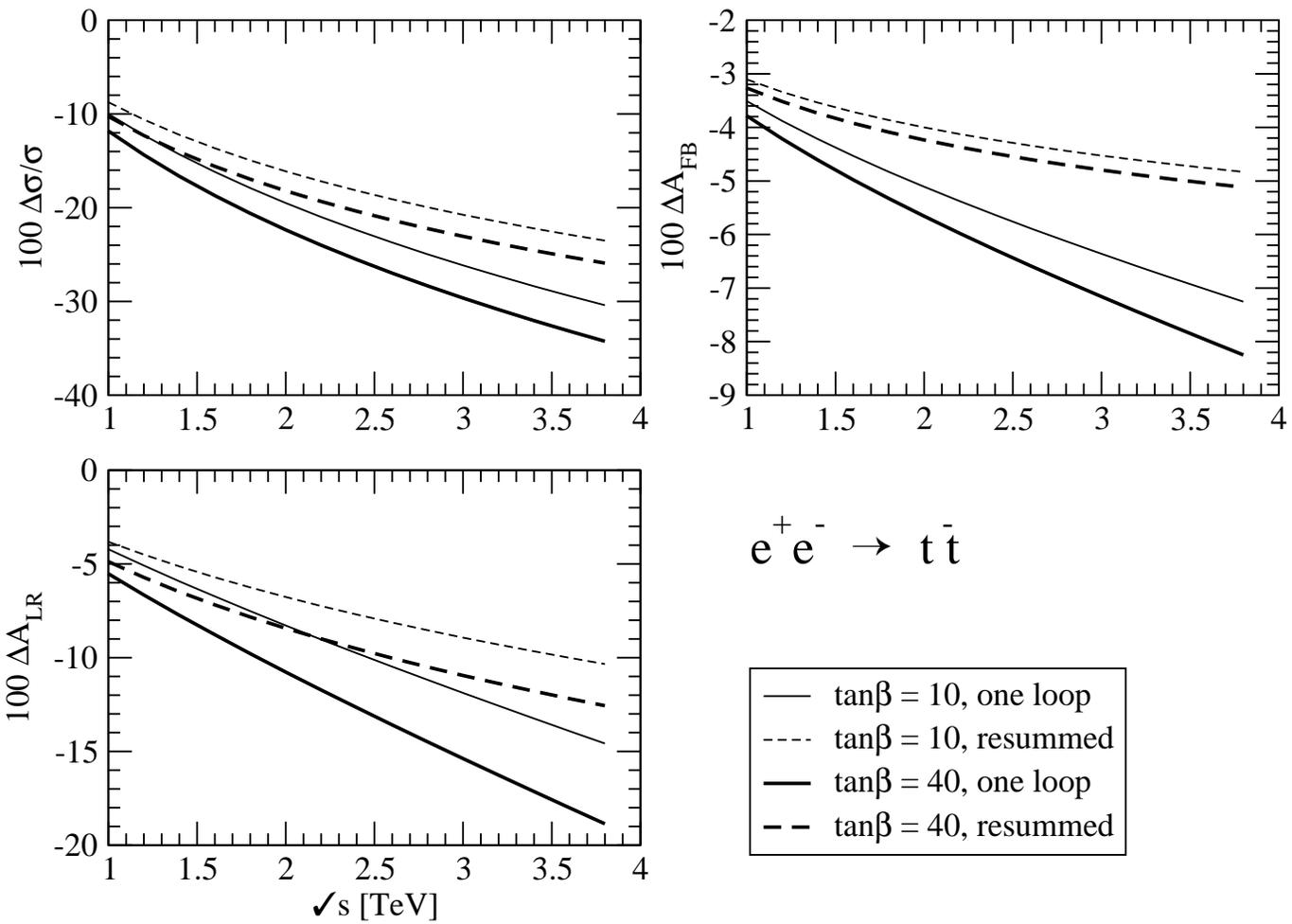}
\vskip-7mm
\end{center}
\caption{Top production. Comparison between the one-loop calculation and the 
resummation at subleading accuracy. 
The mass scales in the logarithms are $M_V=M_W$, $M_{\rm SUSY} = 300$ GeV.}
\label{fig:top}
\end{figure}

\begin{figure}[tb]
\begin{center}
\leavevmode
\epsfig{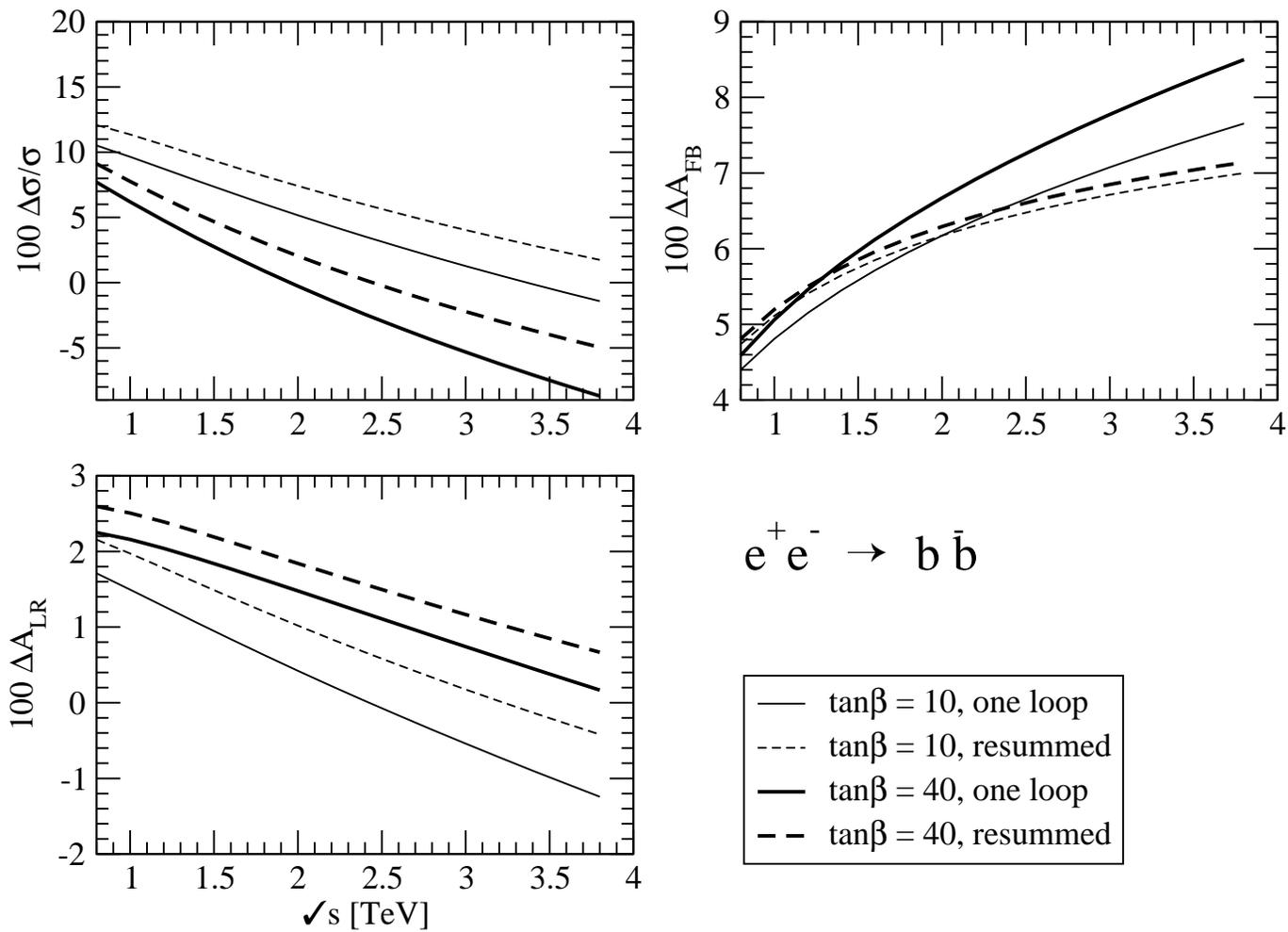}
\vskip-7mm
\end{center}
\caption{Bottom production. Same caption as in Fig.~(\ref{fig:top}).}
\label{fig:bottom}
\end{figure}

\begin{figure}[tb]
\begin{center}
\leavevmode
\epsfig{file=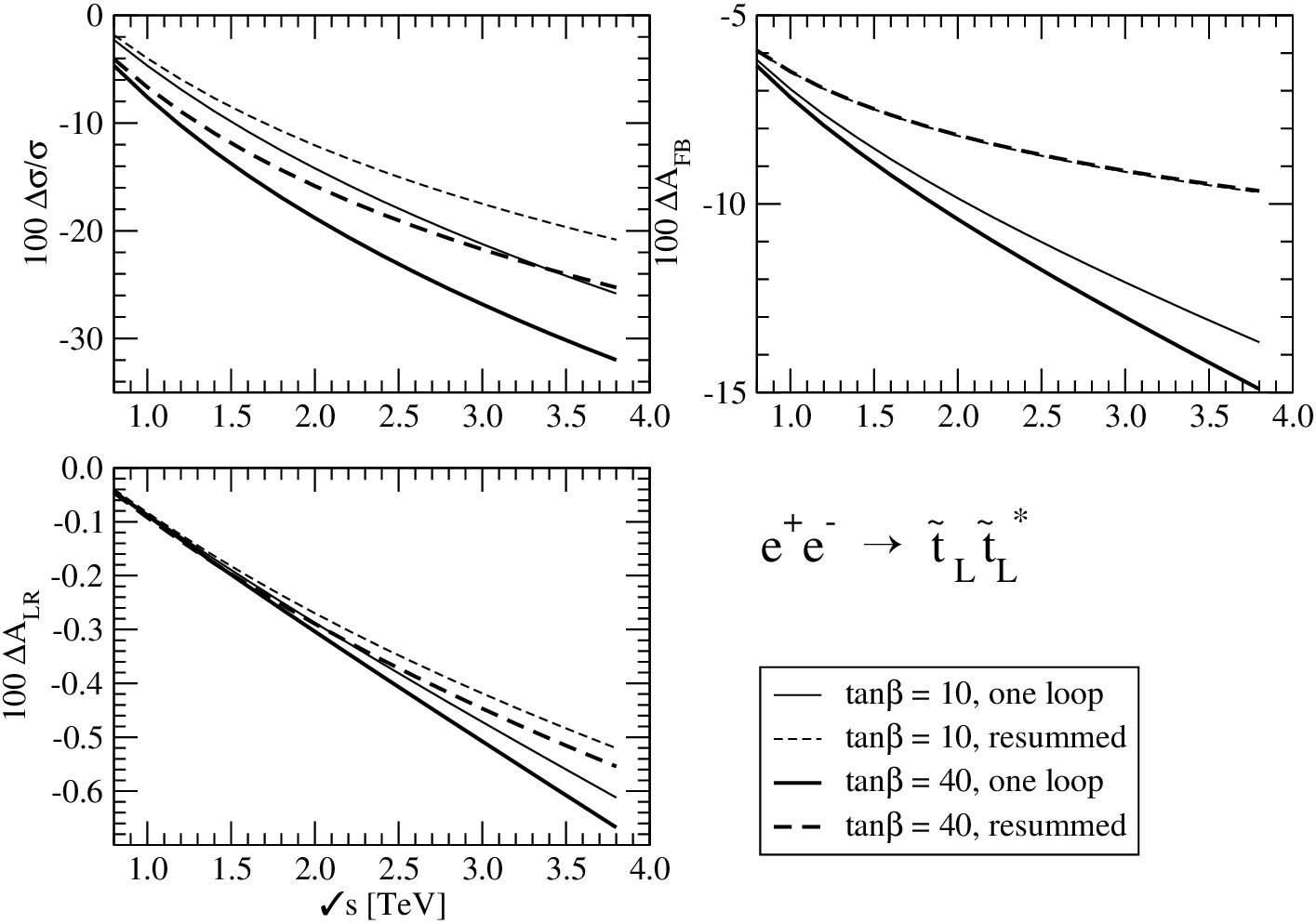,width=1.1\textwidth,angle=90}
\vskip-7mm
\end{center}
\caption{Left handed stop production. Same caption as in Fig.~(\ref{fig:top}).}
\label{fig:stopleft}
\end{figure}

\begin{figure}[tb]
\begin{center}
\leavevmode
\epsfig{file=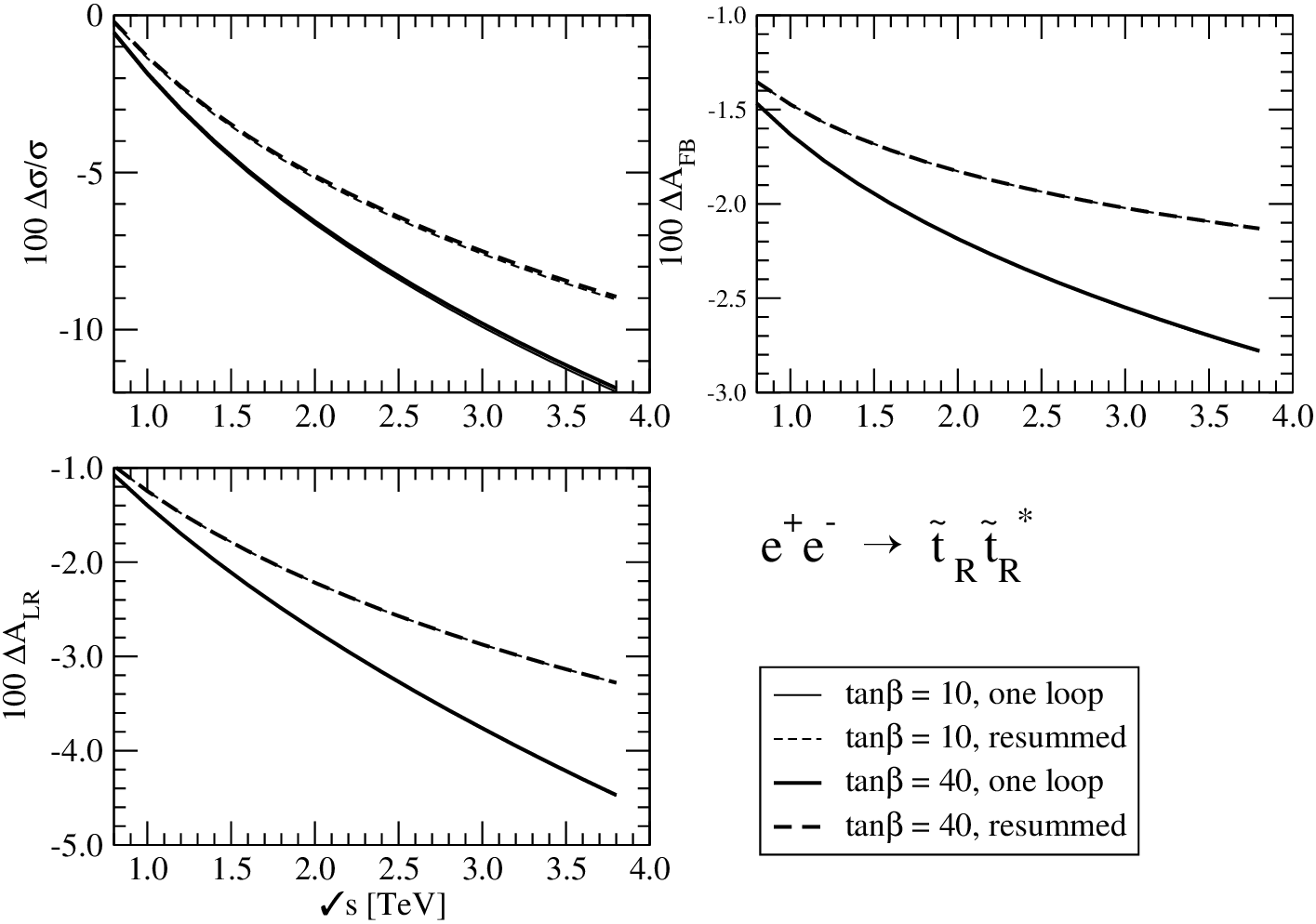,width=1.1\textwidth,angle=90}
\vskip-7mm
\end{center}
\caption{Right handed stop production. Same caption as in Fig.~(\ref{fig:top}).}
\label{fig:stopright}
\end{figure}

\begin{figure}[tb]
\begin{center}
\leavevmode
\epsfig{file=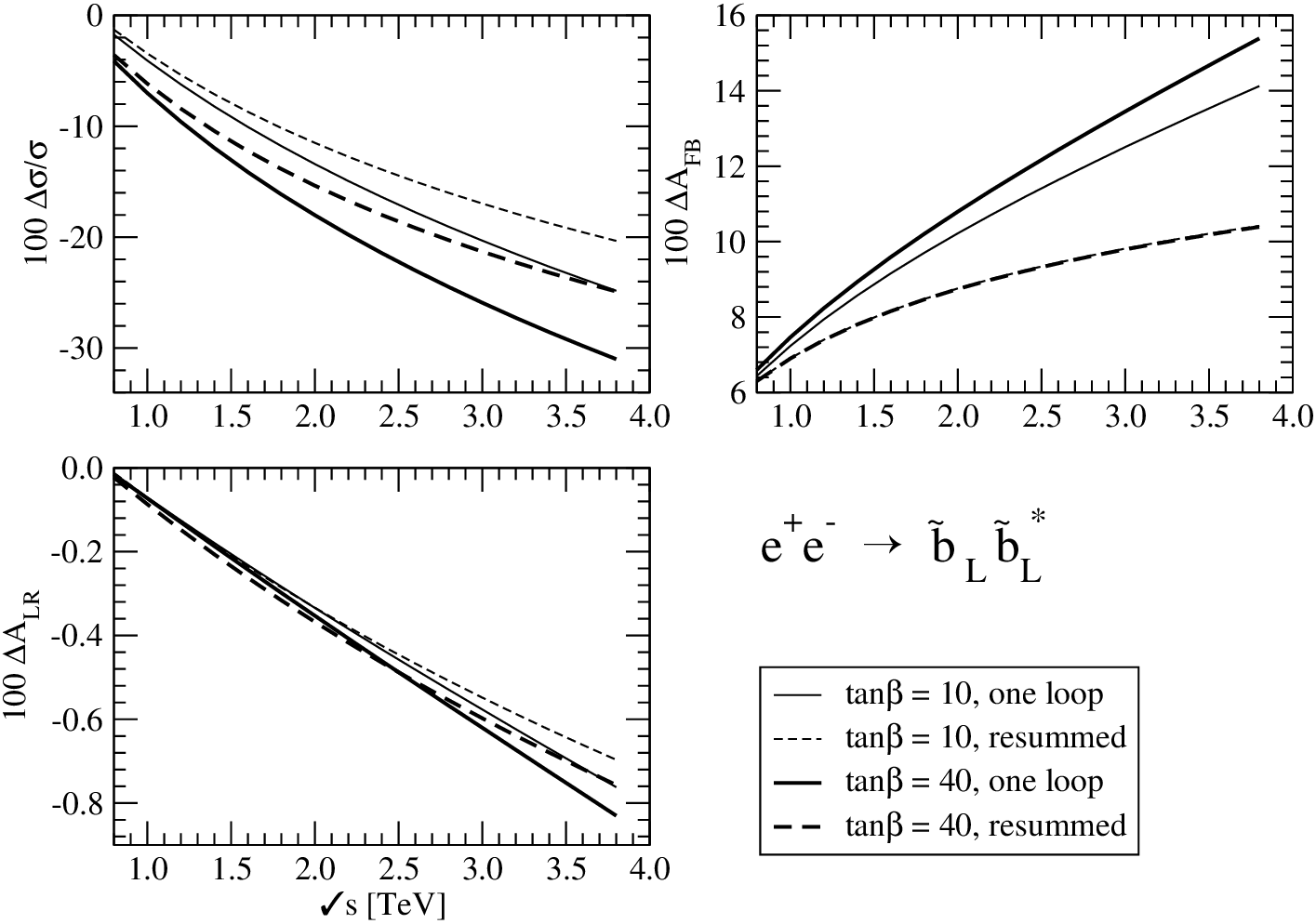,width=1.1\textwidth,angle=90}
\vskip-7mm
\end{center}
\caption{Left handed sbottom production. Same caption as in Fig.~(\ref{fig:top}).}
\label{fig:sbottomleft}
\end{figure}

\begin{figure}[tb]
\begin{center}
\leavevmode
\epsfig{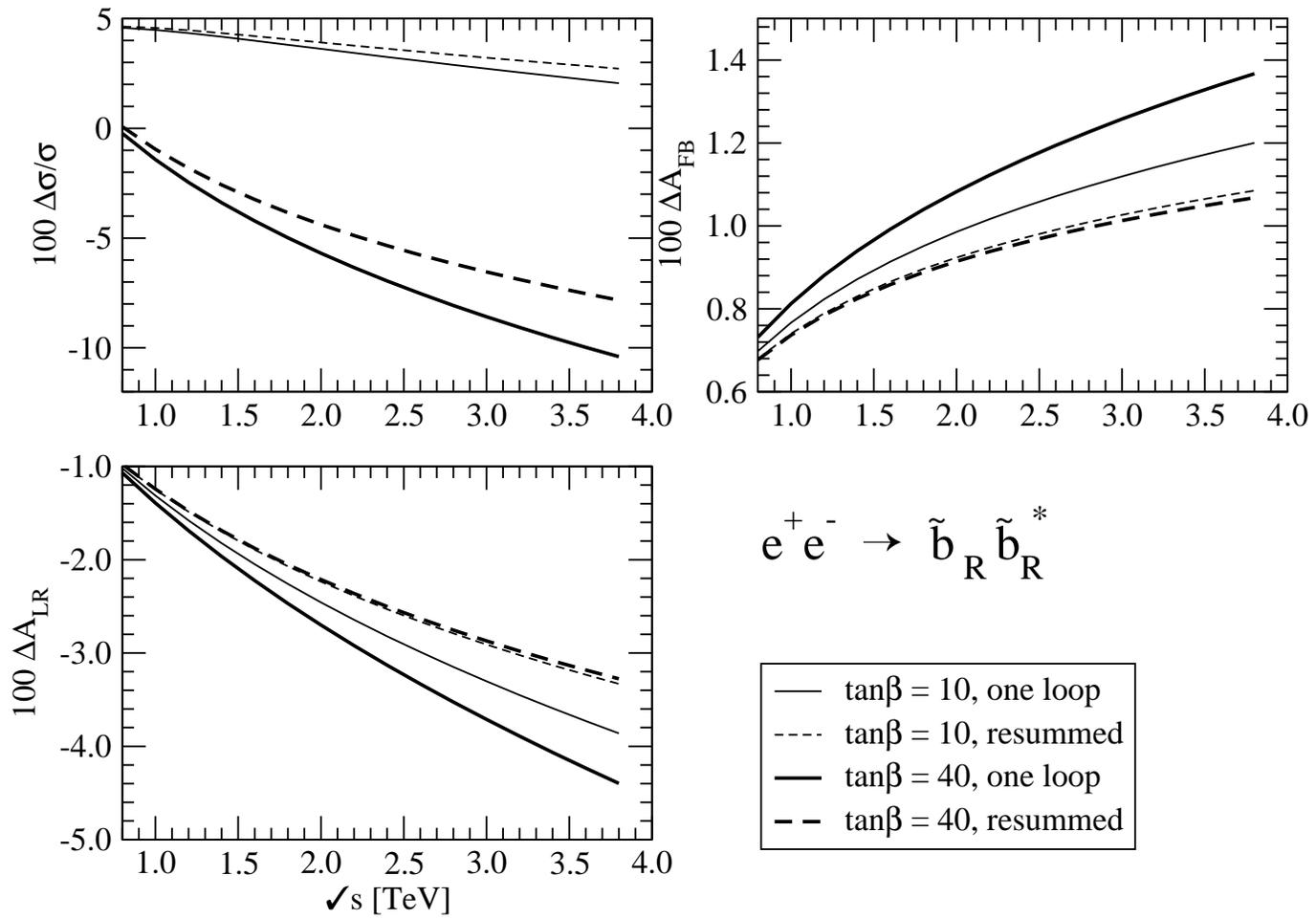}
\vskip-7mm
\end{center}
\caption{Right handed sbottom production. Same caption as in Fig.~(\ref{fig:top}).}
\label{fig:sbottomright}
\end{figure}

\begin{figure}[tb]
\begin{center}
\leavevmode
\epsfig{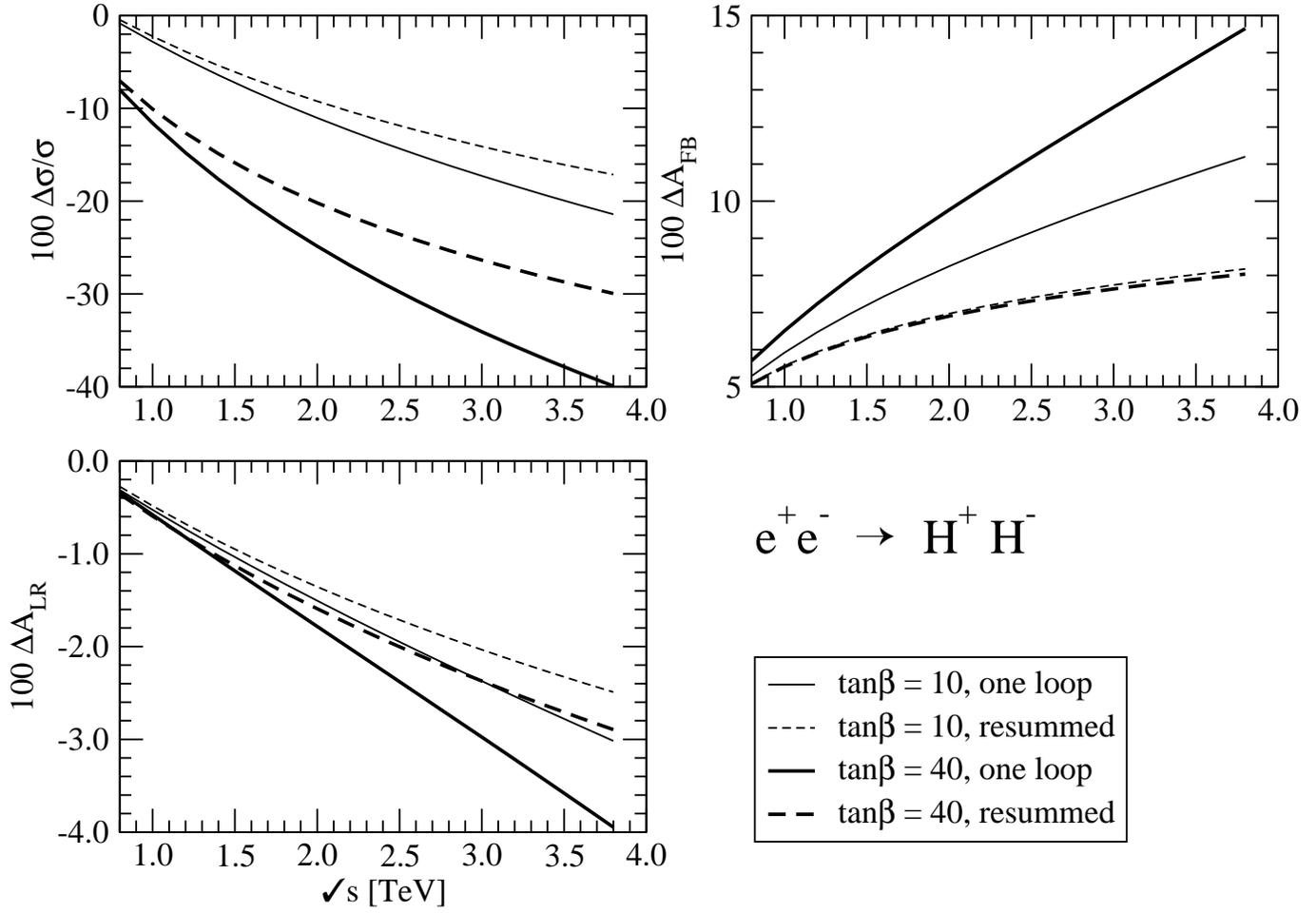}
\vskip-7mm
\end{center}
\caption{Charged Higgs production. Same caption as in Fig.~(\ref{fig:top}).}
\label{fig:higgs}
\end{figure}

\begin{figure}[tb]
\begin{center}
\leavevmode
\epsfig{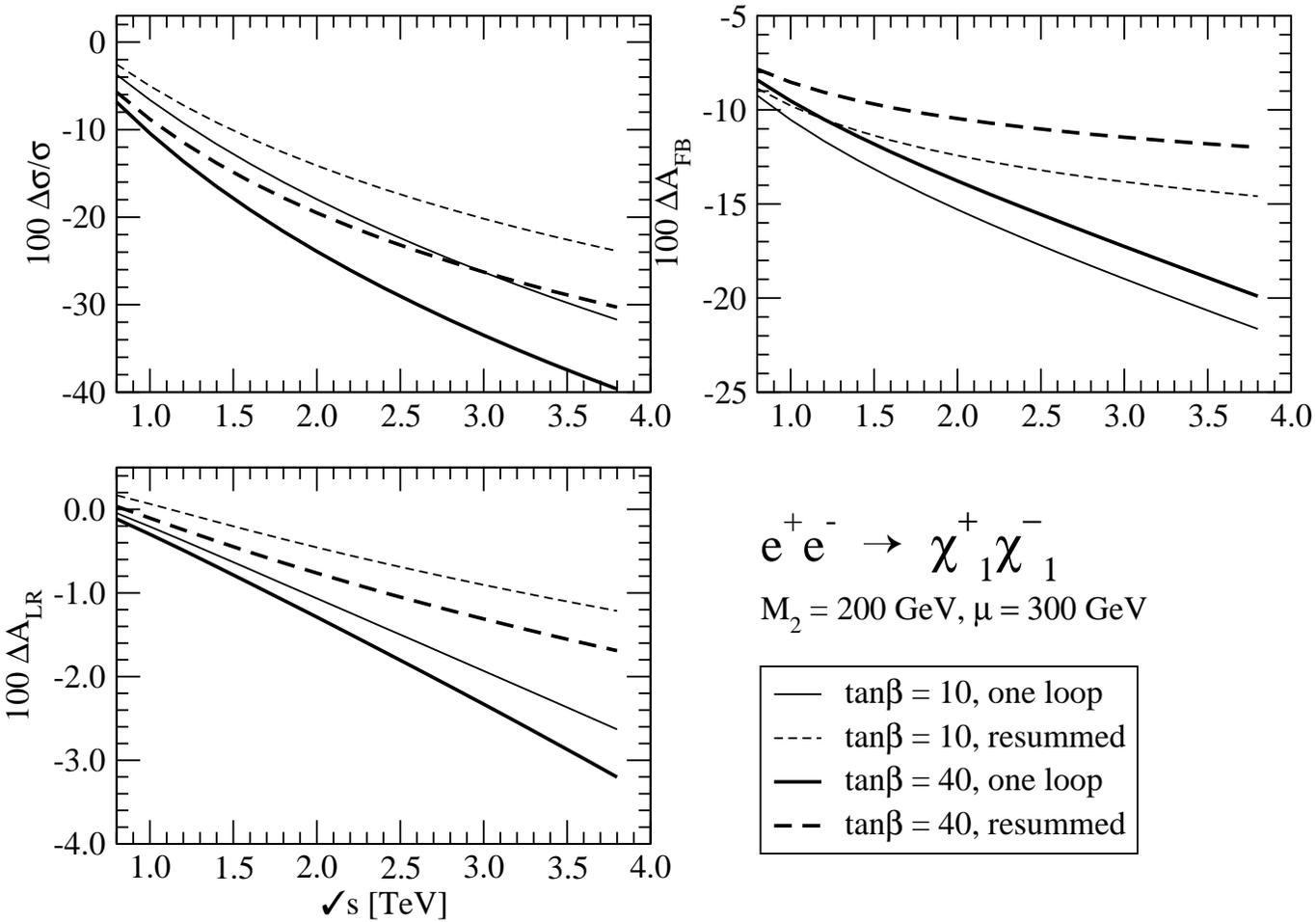}
\vskip-7mm
\end{center}
\caption{Chargino production $e^+e^-\to \chi_1^+\chi_1^-$.
Comparison between the one-loop calculation and the resummation
at subleading accuracy. The mass scales in the logarithms are $M_V=M_W$, $M_{\rm 
SUSY} = 300$ GeV.
The MSSM mixing parameters are $M_2 = 200$ GeV, $\mu = 300$ GeV.}
\label{fig:obs11}
\end{figure}

\begin{figure}[tb]
\begin{center}
\leavevmode
\epsfig{file=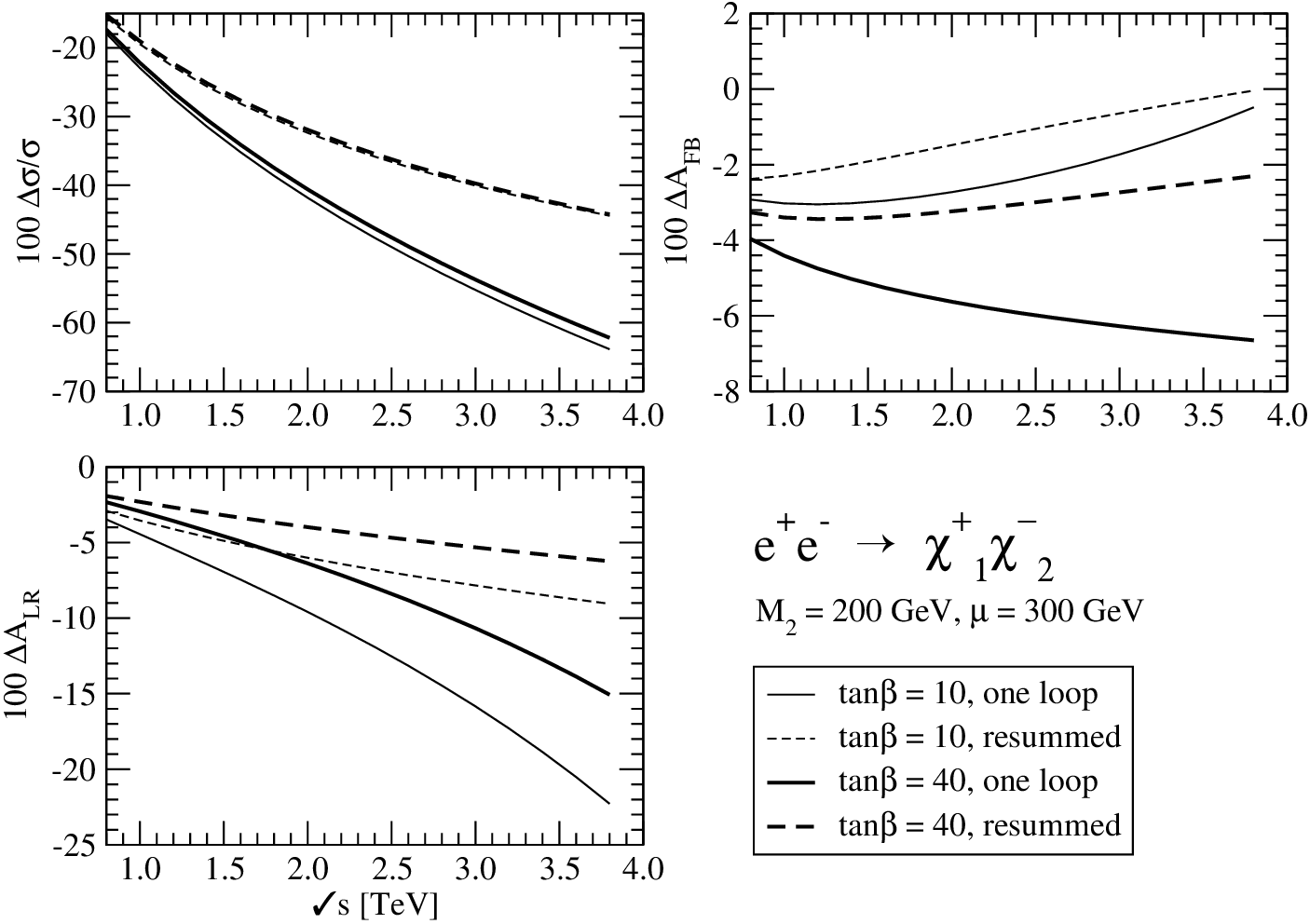,width=1.1\textwidth,angle=90}
\vskip-7mm
\end{center}
\caption{Chargino production $e^+e^-\to \chi_1^+\chi_2^-$.
Same caption as in Fig.~(\ref{fig:obs11}).}
\label{fig:obs12}
\end{figure}

\begin{figure}[tb]
\begin{center}
\leavevmode
\epsfig{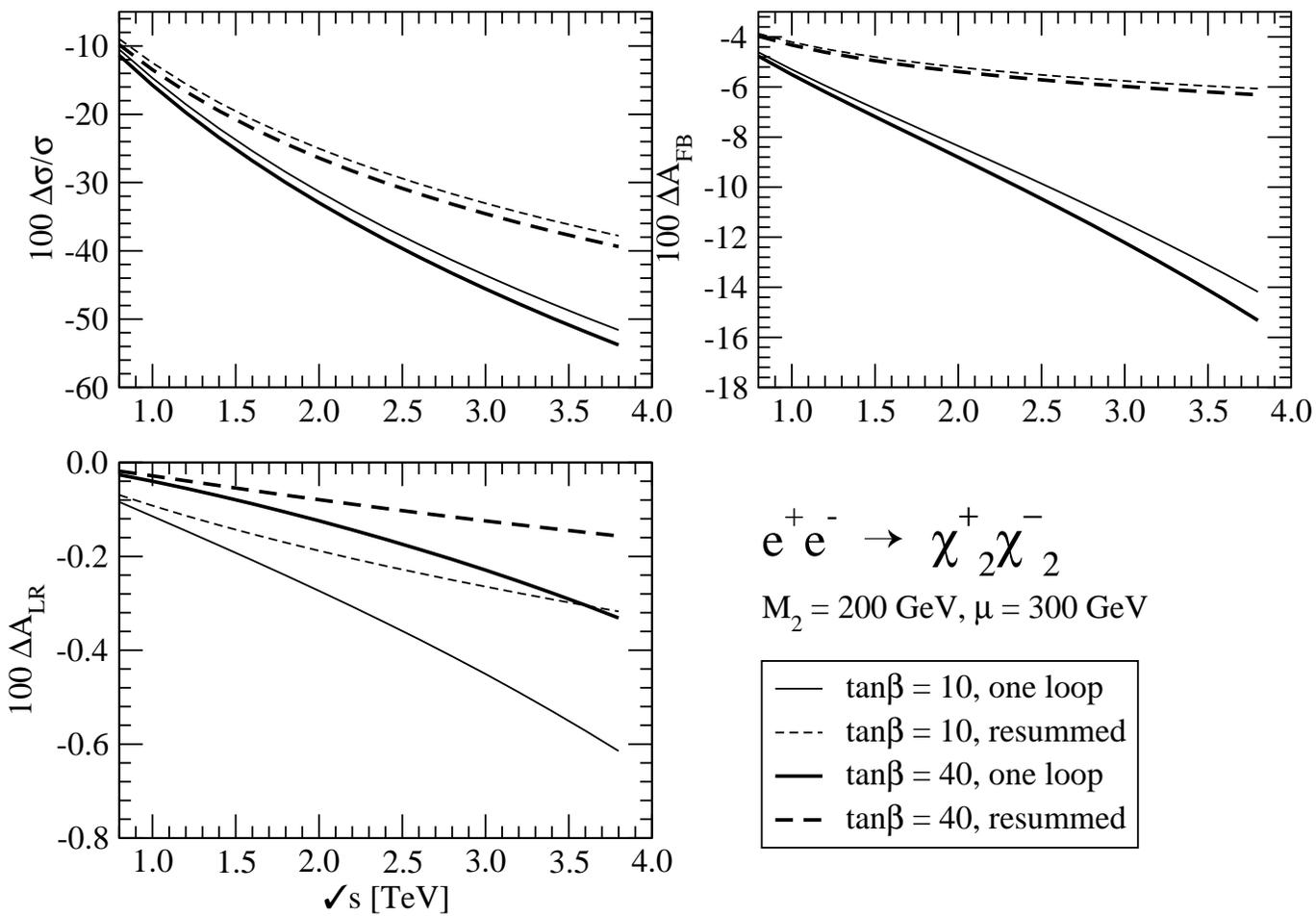}
\vskip-7mm
\end{center}
\caption{Chargino production $e^+e^-\to \chi_2^+\chi_2^-$.
Same caption as in Fig.~(\ref{fig:obs11}).}
\label{fig:obs22}
\end{figure}

\begin{figure}[tb]
\begin{center}
\leavevmode
\epsfig{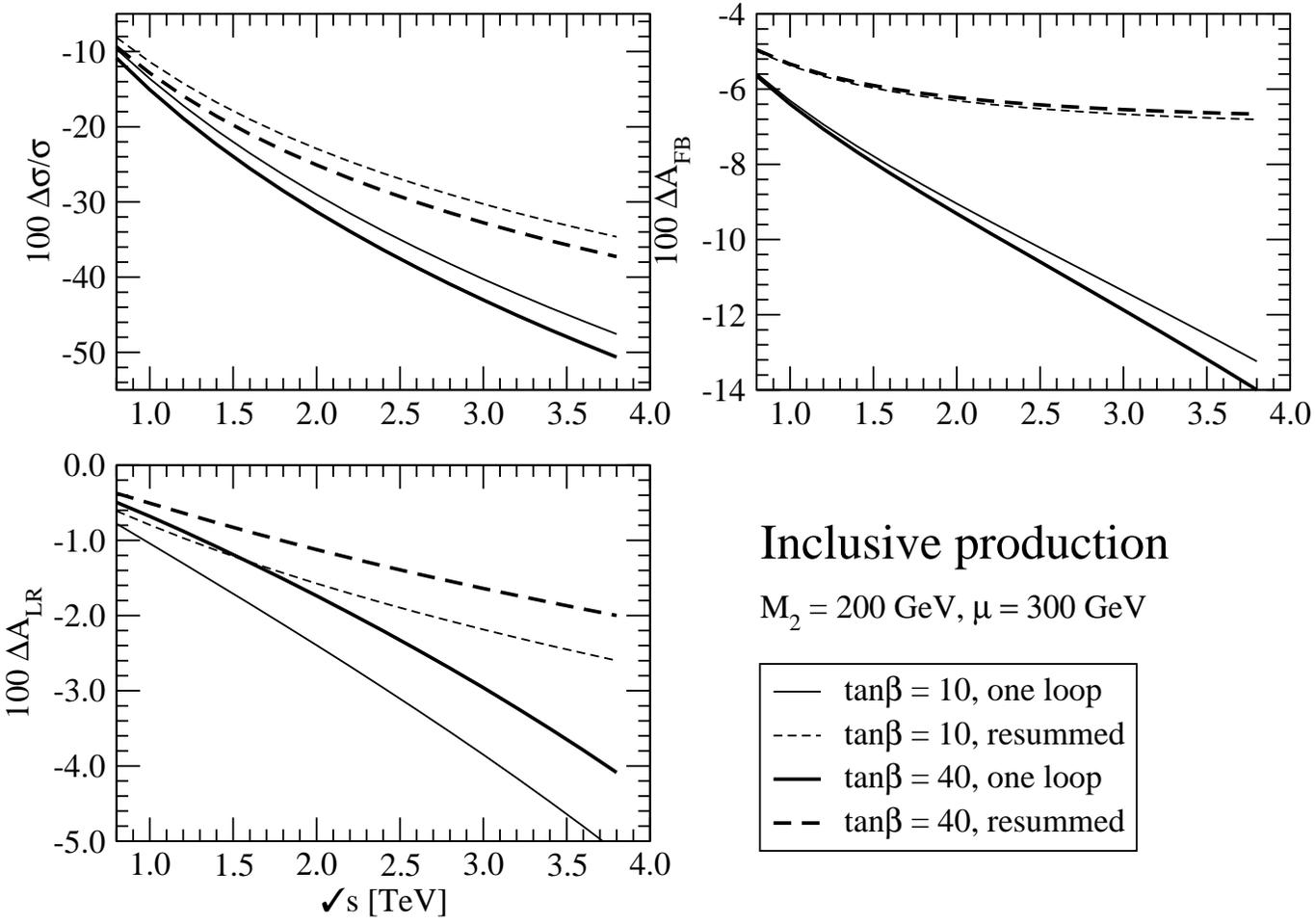}
\vskip-7mm
\end{center}
\caption{Inclusive chargino production (see Eq.(\ref{inclusive})).
Same caption as in Fig.~(\ref{fig:obs11}).}
\label{fig:inclusive}
\end{figure}

\end{document}